\documentstyle[11pt,epsfig]{article}
\def\lsim{\mathrel{\rlap{\lower3pt\hbox{\hskip0pt$\sim$}}
    \raise1pt\hbox{$<$}}}         
\def\gsim{\mathrel{\rlap{\lower4pt\hbox{\hskip1pt$\sim$}}
    \raise1pt\hbox{$>$}}}         
\def\simlt{\mathrel{\raise.3ex\hbox{$<$\kern-.75em\lower1ex\hbox{$\sim$}}}}
\def\simgt{\mathrel{\raise.3ex\hbox{$>$\kern-.75em\lower1ex\hbox{$\sim$}}}}

\begin{document}

\begin{titlepage}
\rightline{May 2004}
\begin{center}

\Large {\bf Effects of Fermion Masses and Twisting on
Non-Integrable Phases on Compact Extra Dimensions}\\[3mm]

\vspace*{0.5cm}
\normalsize

{\bf M{\"u}ge Boz$^{\, a}$ and Nam{\i}k K. Pak$^{\, b}$}

\smallskip
\medskip

{\it $^{a}$ Department of Physics, Hacettepe University, Turkey,
TR06800}

{\it $^{b}$ Department of Physics, Middle East Technical
University, Turkey, TR06531}

\smallskip
\end{center}
\vskip0.2in

\centerline{\large\bf Abstract}

The effective potential for the Wilson loop in the SU(2) gauge
theory with $N_f$ massive fundamental and $N_a$ massive adjoint
fermions on $S^1 \times M^4$ is computed at the one-loop level,
assuming periodic boundary conditions for the gauge field and 
general boundary conditions for fermions. It is
shown that there are critical values for the bare mass, and the
boundary condition parameter for the adjoint fermions, beyond which the
symmetry pattern changes. However,
neither bare mass, nor the boundary condition parameter for the
fundamental fermion play any role on the vacuum structure, thus
the symmetry breaking pattern.
When the two different types of fermions with equal masses exist together 
the pattern of the fundamental fermion dominate, and SU(2)
gauge symmetry remains intact independent of the fermion masses.

\vspace*{2mm}

\end{titlepage}

\section{Introduction}
It has long been known that gauge theories on multiply-connected
spaces exhibit anomalous behavior in that the gauge connection is
promoted to a physical observable. The simplest example is
provided by Aharonov-Bohm effect~\cite{AharonovBohm} according to which the
interference of matter waves in the presence of an impenetrable
domain of magnetic field is modulated by the magnetic flux. This
observation has later been furthered~\cite{Hosotani83} to prove
the dynamical nature of the connection and the irrelevance of
single-valuedness of the matter and gauge fields. The analysis of 
\cite{Hosotani83}, which was focused  on massless fermions in the fundamental
representation of the gauge group, has subsequently been
generalized to adjoint fermions
\cite{Higuchi-Parker,Davies88,Hosotani89}. Recently,
effects of adjoint fermion masses have been incorporated into the
previous works~\cite{Takenaga03}, and it was pointed out that there
exist certain critical values of the fermion masses accross which
the symmetries of the system change.

For appreciating the importance of the Wilson loop dynamics,
consider for definiteness a gauge theory in a 5-dimensional
factorizable geometry $M^{4}\times S^1$ where $S^1$ is a circle
with radius $R$. The gauge field $A_B$ $(B=(\mu,y), \mu=0,1,2,3$)
has five independent components, and it is forbidden to have any
local potential due to higher dimensional gauge invariance.
However, the non-integrable phase factor
\begin{eqnarray}
\theta(x) = - i \ln\left[P e^{i \int_{y=0}^{y=2 \pi R} dy
A_y(x_{\mu},y)}\right]~,
\end{eqnarray}
being inherently non-local in the direction of extra dimension,
develops a non-local potential in the presence of charged bulk
fields. In case $\theta(x)$ develops a nonvanishing vacuum
expectation value (VEV) the gauge symmetry can be broken
dynamically depending on the model parameters~\cite{Hosotani89}. This has been particularly useful in string
compactifications~\cite{Candelas}. Furthermore, recently it has been
pointed out that radiatively-lifted vanishing potential for the
non-integrable phase factor $\theta(x)$ makes it a perfect
candidate for inflaton~\cite{Arkani-Hamed1,Arkani-Hamed2,Riotto} which has to
acquire an extremely flat potential to comply with the
requirements of successful inflation.

In this work,  we consider a non-supersymmetric SU(2) gauge  model
with  $N_f$ massive adjoint fermions, and $N_a$ massive
fundamental fermions, with the most general boundary condition
parameters for the fermions and the gauge fields on $S^1 \times
M^4$. Here is a brief summary of the present work in relation with the 
previous works:

Hosotani has previously considered a SU(2) gauge theory
defined both on  $S^1 \times R^1$ and  $S^1 \times M^3$ with
massless fermions,  but with arbitrary boundary condition (bc)
parameters for the gauge fields, and fermions~\cite{Hosotani89}. He has shown that
the SU(2) gauge symmetry is not broken, when the fermions are in the fundamental representation (FR),
irrespective of the values of the bc 
parameters~\footnote{
In more detail,  for $\delta_f<\pi/2$, absolute minimum is
located at $\theta_m=\pi$, corresponding to 
$U^{sym}=-I$, and  for  $\pi/2< \delta_f<\pi$, the absolute
minimum is located at $\theta_m=0$, corresponding to  $U^{sym}=I$.
Both of these  $U^{sym}$ are elements of the center of SU(2), thus
the SU(2) symmetry is unbroken.}.
But the SU(2) symmetry breaks down to U(1)
 for certain values of the bc parameter below a certain
critical value, for the adjoint representation (AR). Takenaga
more recently considered an SU(2) gauge theory  $S^1 \times M^4$
with massive adjoint fermions, with periodic boundary conditions
for fermions~\cite{Takenaga03}. He has shown that below a certain critical value of
the bare mass the symmetry again breaks down  to U(1).

In Section 3, we considered a SU(2) gauge theory with massive
adjoint fermions and with arbitrary bc parameters. We have shown
that below certain critical values of the bare mass and bc
parameter the symmetry breaks down to U(1), and  agrees with the
results of Hosotani and Takenaga, respectively,   in the corresponding limits.

In Section 4, we extended this discussion 
in~\cite{Hosotani89}
for massless fundamental fermions,
by including bare masses
for fermions. We have shown that neither
bc parameter $\delta_f$, nor the bare
mass for the fundamental fermion play any role on the vacuum
structure/symmetry breaking pattern.

Finally,  in Section 5, we considered the general case with $N_f$
fundamental and $N_a$ adjoint fermions with equal masses. 
We have observed that the fundamental fermions play a 
more dominant role than the adjoint ones, on the gauge symmetry pattern, 
as the result turns out to be very similar to the pure fundamental fermions
case.

\section{The Effective Potential}

Consider an SU(2) gauge theory on $M^4\times S^1$ with $N_a$
adjoint and $N_f$ fundamental fermions. The action is completely
fixed by gauge invariance
\begin{eqnarray}
\label{action}
S=\int d^4x\, dy\left[-\frac{1}{2 g_5^2}\,
tr\left\{F_{A B} F^{A B}\right\}+ \overline{\psi}\left(\gamma^A
D_{A} - m_f\right)\psi + \overline{\lambda}\left(\gamma^A D_A -
m_a \right)\lambda\right]~,
\end{eqnarray}
where $\psi$ and $\lambda$ stand, respectively, for fundamental
and adjoint fermion fields with masses $m_f$ and $m_a$, and $g_5$,
with dimension of (mass)$^{-1/2}$, is the higher dimensional gauge
coupling. The potential for $A_5$ is perfectly flat since gauge
invariance forbids the induction of any local operator which can
lift the flatness. However, the phase of the Wilson loop
$\theta(x)$ is inherently non-local in the extra dimension and
thus it can acquire a non-trivial non-local potential. Indeed, the
gauge field kinetic term in (\ref{action}), after dimensional
reduction, generates the kinetic term
\begin{eqnarray}
{\cal L}_{KK}^{(4)}&=& \frac{1}{2 L^2  g_4^2} \sum_a
(\partial_{\mu} \theta^a )^2~, 
\label{3}
\end{eqnarray}
where  $g_4 = \frac{g_5}{\sqrt {2 \pi R}}$ is the four dimensional
gauge coupling constant, and we  defined a new field such that
\begin{eqnarray}
\theta^{a}(x)= g_5 \  \int_{0}^{2 \pi R} dy \ A^{5,a} (x,y) = 2
\pi R \ g_5  \ A^{5,a} (x)= L g_5  A^{5,a} (x)~, 
\end{eqnarray}
using the compactness of $S^1$ which guarantees the
$y$--independence of the zero mode $A^{5,a}(x,y)$. However, the
exactly flat potential of (\ref{3}) is lifted by the gauge boson
and fermion loops. This radiative contribution, denoted by
$V_{af}$, is given by
\begin{eqnarray}
V_{af}(\theta, \, N_a, \, N_f, \, z_a, \, z_f, \, \delta_a, \, \delta_f) &=&
\frac{ 1}{ c_1 } \Bigg \{ -3
\sum_{n=1}  \frac{1}{n^5} \bigg [1+  \cos 2  n \theta \bigg]\nonumber\\
&+& 
2 \ N_a \sum _{n=1} \frac {F (z_a n)} {n^5} \bigg[ 2 \cos  n
\delta_{a}+ \cos n( 2 \theta+\delta_{a})+\cos n( 2
\theta-\delta_{a})\bigg]\nonumber\\
&+& 2 \ N_f \sum _{n=1} \frac {F (z_f n)} {n^5}  \bigg[ \cos n(
\theta+\delta_{f})+\cos n( \theta-\delta_{f}) \bigg]\Bigg \}~,
\label{5}
\end{eqnarray}
where
\begin{eqnarray}
F (z n)= e^{-z n}  \bigg[1+ z n +\frac{1}{3} z ^2  n^2\bigg] \,\,\,\,\,\,
\mbox{with} \,\,\,\,\,\,
z=m L~,
\end{eqnarray}
and
\begin{eqnarray}
\frac{1}{c_1}= \frac{ 3}{ 2 \pi^2 L^5 }~.
\end{eqnarray}
Here, the $\theta$ is related to $\theta^a$
in Eq. (4), through the 
relationship:
\begin{eqnarray}
L g_5 A^a_5 \tau^a= \theta^a \tau^a= C 
\ \left( \begin{array}{cc}
 \theta   &  0\\ 
 0  &  -\theta   
\end{array}\right)\, C^{+}, 
\end{eqnarray}
with  the constant $2\times 2$ matrix satisfying $C^{+} C=I$, 
and $\theta=\sqrt{\theta_1^2+\theta_2^2+\theta_3^2}$.
We would like to point out, following Hosotani~\cite{Hosotani89},
that because of the invariance of the boundary conditions under global gauge
transformations the effective potential 
does not depend on $C$, and for the $SU(2)$
case depends only on the single $\theta$
and the phases of the fermions.

In  (\ref{5}), the first line follows from the gauge and the ghost fields,
and the second and the third lines are the contributions of $N_a$
massive adjoint and $N_f$ massive fundamental fermions,
respectively. The expression (\ref{5})  reduces correctly to various
special cases already discussed in the literature
\cite{Hosotani83,Higuchi-Parker,Davies88,Hosotani89,Takenaga03}.
One notes that the phases $\delta_a$, and $\delta_f$ are defined
through the boundary conditions. As $S^1$ is not simply connected,
boundary conditions must be specified for the single valuedness of
the observables. For the gauge field
we  adopt periodic boundary conditions; for the fermion fields we impose the
following general boundary conditions:
\begin{eqnarray}
\psi_f (x,y+L)&=&e^{i \delta_f} \psi_f (x,y)~, \nonumber\\
\psi_a (x,y+L)&=&e^{i \delta_a} \psi_a (x,y)~.
\end{eqnarray}
In what follows we discuss the three specific cases of adjoint,
fundamental and adjoint plus fundamental fermions separately. In
each case we analyze the potential landscape both analytically and
numerically with the aim of determining if the original gauge
symmetry is respected by the effective potential~(\ref{5}).

\section{The case with Adjoint Fermions only}

We first consider the case which there are $N_a$ massive adjoint
fermions with the phases $\delta_a$. The effective potential takes
the form
\begin{eqnarray}
V_{a}(\theta, \ N_a, \  z_a, \  \delta_a)&=& \frac{1}{c_1} \sum_{n=1}
\frac{1}{n^5} \bigg[ -3+ 4 \ N_a F (z_a n)  \cos n \delta_a
\bigg] \bigg [1+  \cos 2  n \theta \bigg]~,
\label{apot}
\end{eqnarray}
Note that the effective potential reduces to that of $N_a$
massless adjoint fermions with phases $\delta_a$ when
$z\rightarrow 0$~\cite{Hosotani89}, and differs from the model
considered by Takenaga~\cite{Takenaga03} by the  $\cos n \delta_a$
term multiplying $F_a$.

To identify the role played by the massive fermions with phase
$\delta_a$ on the vacuum structure, we have to look at the two
special limits, namely $z_a \rightarrow \infty$, and $z_a
\rightarrow 0$. The behaviour in the first case is identical to
that of Takenaga~\cite{Takenaga03}, as the fermion is decoupled in
this case. The dominant contribution comes from the gauge sector,
and the vacuum configuration is given by  $\theta=0 \,
[\mbox{mod}\, \pi]$ independent of $\delta_a$. Clearly, the SU(2)
gauge symmetry is not broken in this case.
In the massless limit, $z_a \rightarrow 0$, let us note that
when $\delta_a=0$,  the vacuum configuration
is given by $\theta=\pi/2  \, [\mbox{mod}\, \pi]$~\cite{Davies88}.
However when $\delta_a \neq 0$, we will see that there exists a
critical value $\delta_a^c$, above which $\theta=0 \, [\mbox{mod}
\, \pi]$ is an absolute minimum. To find
the critical value $\delta_c^{(a)}$, we define:
\begin{eqnarray}
c_2  V_{a}^{\prime \prime }(\theta=0 \, [\mbox{mod} \, \pi], \ N_a, \ z_a=0, 
\ \delta_a)&=& \sum_{n=1} \frac{1}{n^3} \bigg [3 - 4 \ N_a
\  \cos n \ \delta_a \bigg]~,
\end{eqnarray}
where
\begin{eqnarray}
c_2&=& \frac{c_1}{4}~,
\end{eqnarray}
and we have used the standard definitions:
\begin{eqnarray}
\sum \frac{\cos n \pi}{n^D}&=&-\bigg[1-2^{(-D)}\bigg] \xi_D, \,\,\, \mbox{and}\nonumber\\
\xi_D &=&\sum \frac{1}{n^D}~,
\end{eqnarray}
We  plot $c_2 V_{a}^{\prime\prime} (\theta= 0 \,
[\mbox{\small{mod}} \, \pi], \, N_a=1, \,  z_a=0, \, \delta_a)$ with respect
to  $\delta_a$ in Figure 1. As can be seen from Figure 1 
that there is a critical value at $\delta_{a}^{c_{1}}=0.53$.
\begin{figure}[htb]
\vskip -2.8truein 
\centering \epsfxsize=5.5in
\leavevmode\epsffile{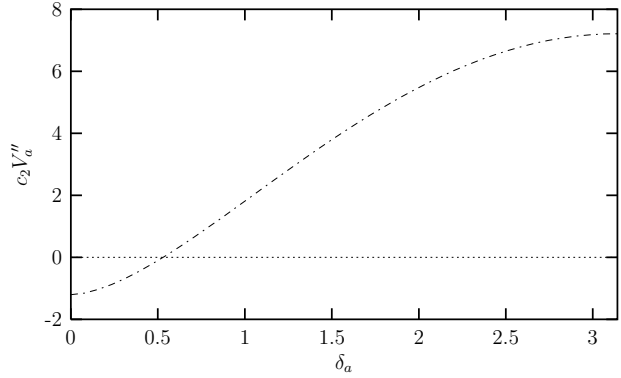}
\vskip -3.15truein 
\caption[]{The
$\delta_a$ dependence of $c_2  V_{a}^{\prime\prime} (\theta= 0 \, [\mbox{\small{mod}} \, \pi], \,
N_a=1, \  z_a=0, \, \delta_a)$.} 
\label{fig1}
\end{figure}

Consistency with the results of Davies and McLachan~\cite{Davies88} requires
that there must be a  critical value  (same or different than $\delta_a^{c_{1}}$)  
below which $\theta=\pi/2 \, [\mbox{\small{mod}} \, \pi]$ is an absolute
minimum. To find this critical value, we again define
\begin{eqnarray}
c_2  V_{a}^{\prime \prime }(\theta=\pi/2 \, [\mbox{\small{mod}} \,
\pi], \ N_a, \   z_a=0, \ \delta_a) &=& \sum_{n=1}  \frac{(-1)^n}{n^3}
\bigg[ 3- 4 \ N_a  \cos n \delta_a\bigg]~,
\end{eqnarray}
and in Figure 2, we investigate  the dependence   of $c_2 V_{a}^{\prime \prime } (\theta=\pi/2 \, [\mbox{\small{mod}} \,
\pi], \ N_a=1,  \  z_a=0, \ \delta_a)$ on  $\delta_a$. 
\begin{figure}[htb]
\vskip -2.8truein 
\centering \epsfxsize=5.5in
\leavevmode\epsffile{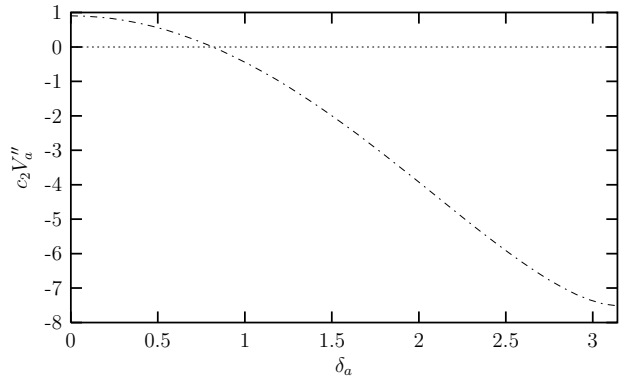}
\vskip -3.15truein 
\caption[]{The
$\delta_a$ dependence of $c_2  V_{a}^{\prime\prime} (\theta= \pi/2
\, [\mbox{\small{mod}} \, \pi],  \, N_a=1, \,   z_a=0, \, \delta_a)$.} 
\label{fig2}
\end{figure}

As can
be seen from Figure 2 that there is a critical point at
$\delta_a^{c_{2}}=0.81$, which is different from the previous
case. 

If we summarize,

$(i)$ For  $\delta_{a}^{c_{2}}< \delta < \pi, \,\,\, \theta=0
\,\,[\mbox {mod} \,\, \pi]$ is an absolute  minimum, and the SU(2)
gauge symmetry is not broken.

$(ii)$ For  $0< \delta_a<  \delta_{a}^{c_{1}}, \, \,\,\theta=\pi/2 \, \, [\mbox {mod} \,\, \pi]$ is an absolute
minimum, and the gauge symmetry is dynamically broken down to
U(1).

This observation suggests that there must exist certain critical
values of $z_a$ at which gauge symmetry breaking patterns change
when  $0 <   \delta_a <  \delta_{a}^{c_{1}}$ (note that this does
not happen  when   $ \delta_{a}^{c_{2}} <  \delta_a < \pi$, as
there is no difference in the symmetry breaking structure from
$z_a \rightarrow \infty$ to  $z_a \rightarrow 0$).

Before adressing the stability question of the vacuum
configurations identified above, we would like to study the
interval $\delta_{a}^{c_{1}} < \delta_a <  \delta_{a}^{c_{2}}$ in
detail. Figure 1 and Figure 2 suggest that in this interval of $
\delta_a$ all the three vacuum configurations, namely $\theta=0,\
\pi/2$ exist simultaneously. The behaviour of $c_1
V_{a}(\theta, \, N_a=1, \ z_a=0, \,  \delta_a)$ for different values of
$\delta_a$ in this interval is plotted in Figure 3. 
\begin{figure}[htb]
\vskip -2.7truein 
\centering \epsfxsize=5.5in
\leavevmode\epsffile{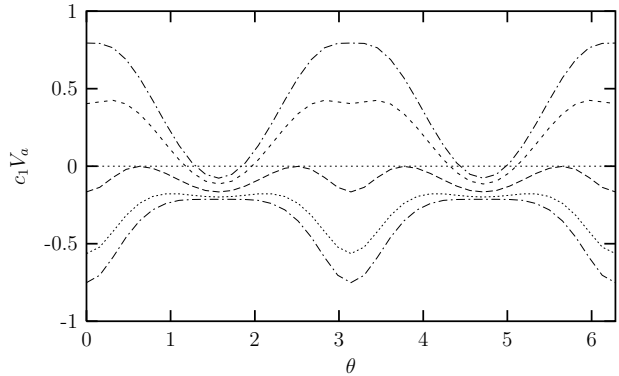}
\vskip -3.15truein
\caption[]{The dependence on $\theta$  of $ c_1 V_{a}  
(\theta, \, N_a=1, \, z_a=0, \,  \delta_a)$. Here, the top and
bottom dot-dashed curves represent $\delta_a^{c_{1}}=0.53$, and
$\delta_a^{c_{2}}=0.81$ values of $\delta_a$ respectively, whereas
the three different curves in the $\delta_{a}^{c_{1}} < \delta_a<
\delta_{a}^{c_{2}}$ interval correspond to $\delta_a= 0.61, \  0.71,\ 
0.78$, from top to bottom, respectively.} 
\label{fig3}
\end{figure}

We see from Figure 3 that for $\delta_a $ close to
$\delta_{a}^{c_{2}}$, \, $\theta=0 [\mbox {mod} \,\, \pi]$ and
$\delta_a$ close to  $\delta_{a}^{c_{1}}$, \, $\theta=\pi/2 [\mbox
{mod} \,\, \pi]$ are  the absolute minima, respectively. Somewhere
in between, namely at $\delta_a=0.71$, the two are degenerate.
That is in the massless case, for the
values of $\delta_a$ within the interval   $(\delta_{a}^{c_{1}},
\,\, \delta_{a}^{c_{2}})$,
there exists a mixed phase, namely unbroken SU(2) phase
together with the broken phase U(1). This interesting phenomena clearly
deserves further study, which we postpone to a future work.

To confirm the existence of the critical values of $z_a$ we have to
study the stability of the configurations $\theta=0$~[mod $\pi$],
and  and $\theta=\pi/2$~[mod $\pi$] with respect to $z_a$,
corresponding to vacuum configurations, in the limits $z_a \rightarrow \infty$ as well as $z_a \rightarrow 0$ (when
$\delta_{a}^{c_{2}} <\delta_a  < \pi$), and  $z_a \rightarrow 0$
(when $0< \delta_a  <\delta_{a}^{c_{1}}$), respectively.

The second derivative of 
the effective potential 
is plotted with
respect to $z_a$ for $\theta=0$[mod $\pi$] in Figure 4, and for
$\theta=\pi/2$ [mod $\pi$] in Figure 5, for $N_a=1$, 
with their explicit expressions given as : 
\begin{eqnarray}
c_2 V_{a}^{\prime \prime }(\theta=0  \, [\mbox {mod} \, \pi], \ N_a, \  z_a, \ \delta_{a}^{c_{2}}  <\delta_a < \pi )&=&
 \ 3 \xi_3 - 4 \ N_a   \sum_{n=1}\frac{F(z_a n) \cos n
\delta_a}{n^3}~,\nonumber\\
c_2 V_{a}^{\prime \prime }(\theta=\pi/2 \, [\mbox {mod} \, \pi],
\, N_a,  \, z_a, \, 0 < \delta_a < \delta_{a}^{c_{1}})&=&
-\frac{9}{4} \xi_3  -  4 \ N_a \sum_{n=1}\frac{(-1)^n F (z_a n) \cos n
\delta_a }{n^3}~,
\end{eqnarray}
\begin{figure}[htb]
\vskip -2.7truein 
\centering \epsfxsize=5.5in
\leavevmode\epsffile{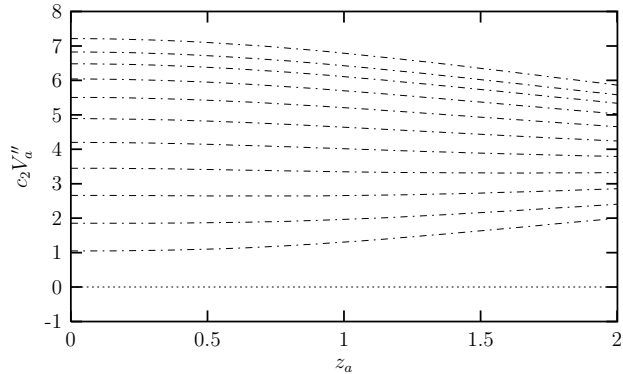}
\vskip -3.15truein
\caption[]{The $z_a$ dependence of $ c_2 V_{a}^{\prime\prime}
(\theta=0 \, [\mbox {mod} \,\, \pi], \ N_a=1, \  z_a, \
\delta_{a}^{c_{1}} <\delta_a<\pi)$. Here,  $\delta_{a}^{c_{2}}=0.81$ for the bottom curve, whereas
$\delta_a=\pi$ for the top curve.} 
\label{fig4}
\end{figure}

We see from Figure 4 that $c_2 V_{a}^{\prime \prime }(\theta=0, \ N_a=1, \
z_a, \ \delta_a^{c_{2}} <\delta < \pi )$ is always positive, and
there is no critical value $z_a^c$ where $V_{a}^{\prime \prime }$
changes sign, and $\theta=0 \, [\mbox{mod} \ \pi]$ is stable
independent of $z_a$. This is consistent with the previous
observation that as long as   $\delta_a^{c_{2}}<\delta_a < \pi$,
 $ \theta=0 \, [\mbox {mod} \pi]$
is an absolute minimum both in
 $z_a \rightarrow \infty$ and  $z_a \rightarrow 0$
limits.
\begin{figure}[htb]
\vskip -2.7truein 
\centering \epsfxsize=5.5in
\leavevmode\epsffile{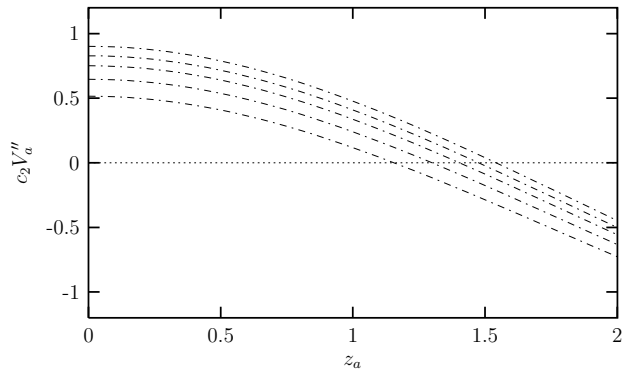}
\vskip -3.15truein
\caption[]{The $z_a$ dependence of $ c_2 V_{a}^{\prime\prime}
(\theta=\pi/2 \, [\mbox {mod} \,\, \pi], \ N_a=1, \  z_a,  \  0< \delta <
\delta_{a}^{c_{1}})$. Here,  $\delta_a=0$ for the 
top curve, whereas $\delta_a^{c_{1}}=0.53$ for the bottom curve.} 
\label{fig5}
\end{figure}

We see from  Figure 5 that there are critical values for $z_a$,
depending on the values of $\delta_a$, below which $V_{a}^{\prime
\prime }(\theta=\pi/2, \  N_a=1, \  z_a, \  0 < \delta < \delta_a^{c_{1}})$
is positive. The largest of these $z_a^c$ corresponding to
$\delta_a=0$ is  $z_a^c=1.5$, which is identical to the result of
Takenaga~\cite{Takenaga03}. The larger $\delta_a$ is, within the
allowed range (0, \ $\delta_{a}^{c_{1}}$), the smaller $z_a$ gets.
That is,  the symmetry breaking pattern is more sensitive to
adjoint mass for larger values of the  phase $\delta_a$, in the
allowed range   $(0, \ \delta_a^{c_{1}})$.

If we summarize

$(i)$ when $0 < \delta_a <  \delta_a^{c_{1}} $,\,\,
$V_a^{\prime\prime}(\theta=\pi/2, \ N_a=1, \  z_a, \  \delta_a)>0$ for a set of
values for $z_a<z_a^c=1.5$, the gauge symmetry is broken to U(1),

$(ii)$ when $\delta_a^{c_{2}}  < \delta_a < \pi $,\,\,
$V_a^{\prime\prime}(\theta=0 \,[\mbox{mod} \pi], \ N_a=1, \  z_a, \ \delta_a)>0$,
independent of the values of $z_a$, and the gauge symmetry SU(2)
is intact.

$(iii)$ when $\delta_a^{c_{2}}  < \delta_a < \delta_a^{c_{1}} $,
there exists a mixed phase, namely the unbroken SU(2) phase
together with the broken phase U(1).
\begin{figure}[htb]
\vskip -2.7truein 
\centering \epsfxsize=5.5in
\leavevmode\epsffile{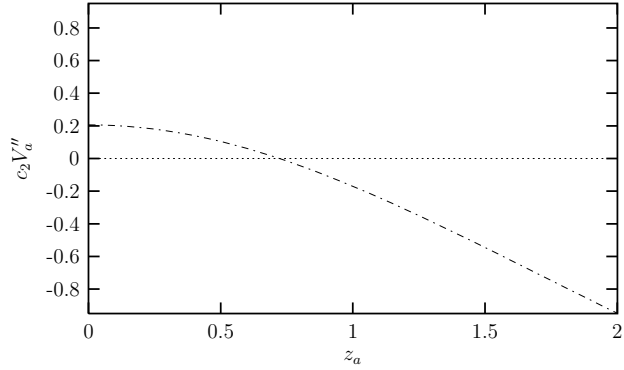} 
\vskip -3.15truein 
\caption[]{The
$z_a$ dependence of $ c_2 V_{a}^{\prime\prime} (\theta=\pi/2 \,
[\mbox {mod} \,\, \pi], \  N_a=1, \ z_a, \   \delta_a=0.71 )$.}
\label{fig6}
\end{figure}

For instance, in Figure 6, we show the  $z_a$ dependence of $ c_2
V_{a}^{\prime\prime} (\theta=\pi/2 \, [\mbox {mod} \,\, \pi], \, N_a=1, \ z_a, \, \delta_a)$
when    $\delta_a=0.71$, for which case the two
minima are degenerate (see Figure 3),
with the critical value $z_c=0.72$. 
We would have obtained identical information if we have plotted 
$ c_2 V_{a}^{\prime\prime} (\theta=0 \, [\mbox {mod} \,\, \pi], \  N_a=1, \
z_a, \ \delta_a=0.71)$.

\section{The Case with Fundamental fermions only}

When there are $N_f$ massive fundamental fermions only,  with the
bc phases  $\delta_f$,  the potential takes the
form:

\begin{eqnarray}
V_{f}(\theta, \ N_f, \  z_f, \ \delta_f)&=&\frac{1}{c_1}
\sum_{n=1}\frac{1}{n^5} \bigg[ -3 (1+ \cos 2  n \theta)  + \ 4 \
N_f F (z_f n) \  \cos n \delta_{f}  \cos n  \theta\bigg]~.
\label{fpot}
\end{eqnarray}

The effective potential reduces to that
of $N_f$ massless fundamental fermions with phases
$\delta_f$~\cite{Hosotani83,Hosotani89} in the $z_f \rightarrow 0$
limit. Again, to identify the role played by the masses and the
$\delta_f$-phases of the fermion on the vacuum structure, we have
to look at the $z_f  \rightarrow \infty$, and $z_f  \rightarrow 0$
limits. As the fermions decouple in the former case, this case is
identical to that of adjoint fermions, and the vacuum structure is
given by $\theta=0 \,[\mbox{mod}\, \pi]$ independently of
$\delta_f$. The SU(2) gauge symmetry is intact in this regime.

Next, we look at the  $z_f  \rightarrow 0$ limit, in
detail. First recall that this limit with $\delta_f \neq 0$ was
considered by Hosotani~\cite{Hosotani83}, for $M^3\times S^1$. He
has  shown that the absolute minimum is $\theta=0$, for $\pi/2 <
\delta_f <\pi$, and $\theta=\pi$ for  $0 < \delta_f <\pi/2$
independent of the number of fermions $N_f$; furthermore these two
absolute minima are degenerate for $\delta_f=\pi/2$. He
has further shown, as mentioned before in the footnote, that in both cases the gauge
symmetry is unbroken. We have $M^4\times S^1$; therefore the
critical value of $\delta_f$ ($\delta_f^c$), if there is any,
could be different than that of Hosotani~\cite{Hosotani83}. We
first plot the expressions  for   $V_{f}^{\prime\prime
}(\theta=0, \, N_f, \, z_f=0,\, \delta_f)$, and   $V_{f}^{\prime\prime }(\theta=\pi,
\, N_f, \ z_f=0, \, \delta_f)$, to
identify the $\delta_f$-regions where $\theta_m=0,\, \pi$ are the  minima,
respectively.

Using  (\ref{fpot}), we  get:

\begin{eqnarray}
c_2 V_{f}^{\prime\prime }(\theta=0, \, N_f, \, z_f=0,\, \delta_f)&=&
\sum_{n=1}\frac{1}{n^3} \bigg[3- \ N_f  \cos n \delta_{f}\bigg]~,\nonumber\\
c_2 V_{f}^{\prime\prime }(\theta=\pi, \ N_f, \  z_f=0, \ \delta_f)&=&
\sum_{n=1}\frac{1}{n^3} \bigg[ 3 -\ N_f   (-1)^n \cos n
\delta_{f}\bigg]~,
\end{eqnarray}
and analyze the  dependence on $\delta_f$  of $c_2
V_f^{\prime\prime} (\theta=0, \ N_f, \  z_f=0, \ \delta_f )$ and $c_2
V_f^{\prime\prime} (\theta=\pi, \ N_f, \  z_f=0, \ \delta_f )$ in Figure 7
and Figure 8, respectively   when $N_f=1,\ 2,\ 3,\ 4$.
\begin{figure}[htb]
\vskip -2.8truein 
\centering \epsfxsize=5.5in
\leavevmode\epsffile{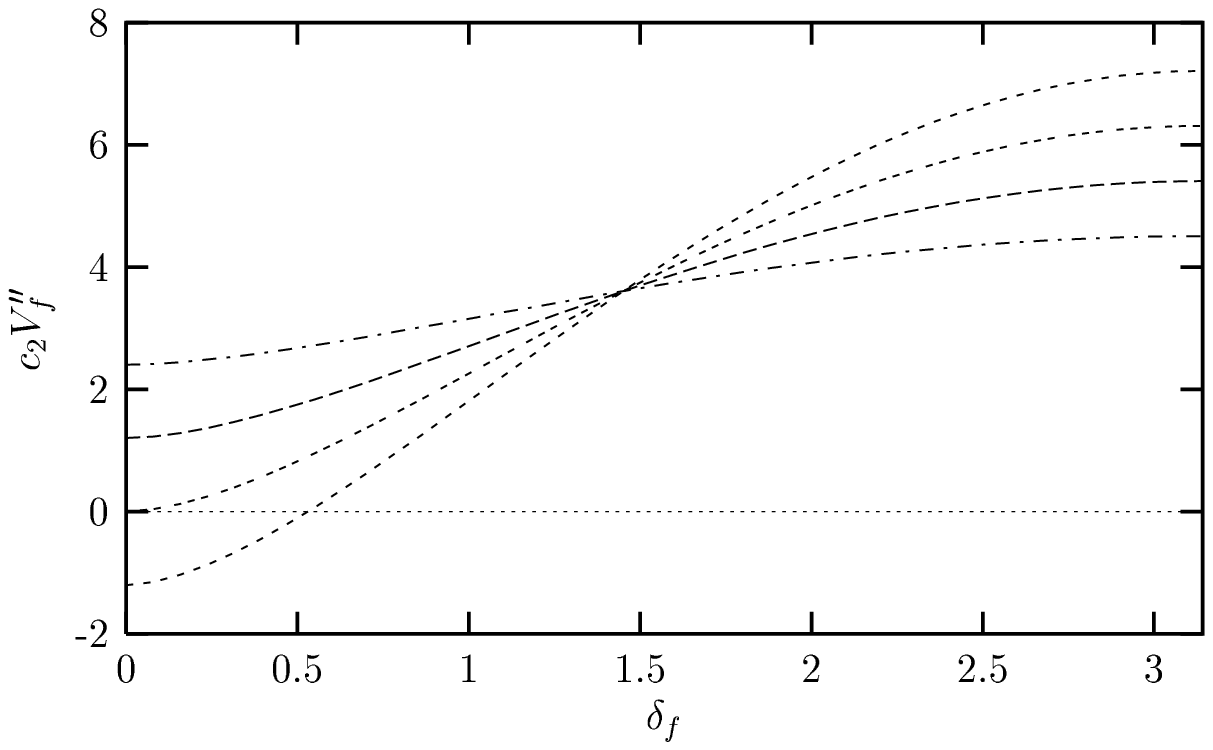}
\vskip -3.15truein 
\caption[]{The
$\delta_f$ dependence of $c_2 V_f^{\prime\prime} (\theta=0, \  N_f, \ z_f=0,
\ \delta_f )$, when $N_f=1$ (top curve),  and   $N_f=4$
(bottom curve). The curves in between are  for $N_f=2$, \  $N_f=3$,
from top to bottom.} 
\label{fig7}
\end{figure}
\begin{figure}[htb]
\vskip -2.7truein 
\centering \epsfxsize=5.5in
\leavevmode\epsffile{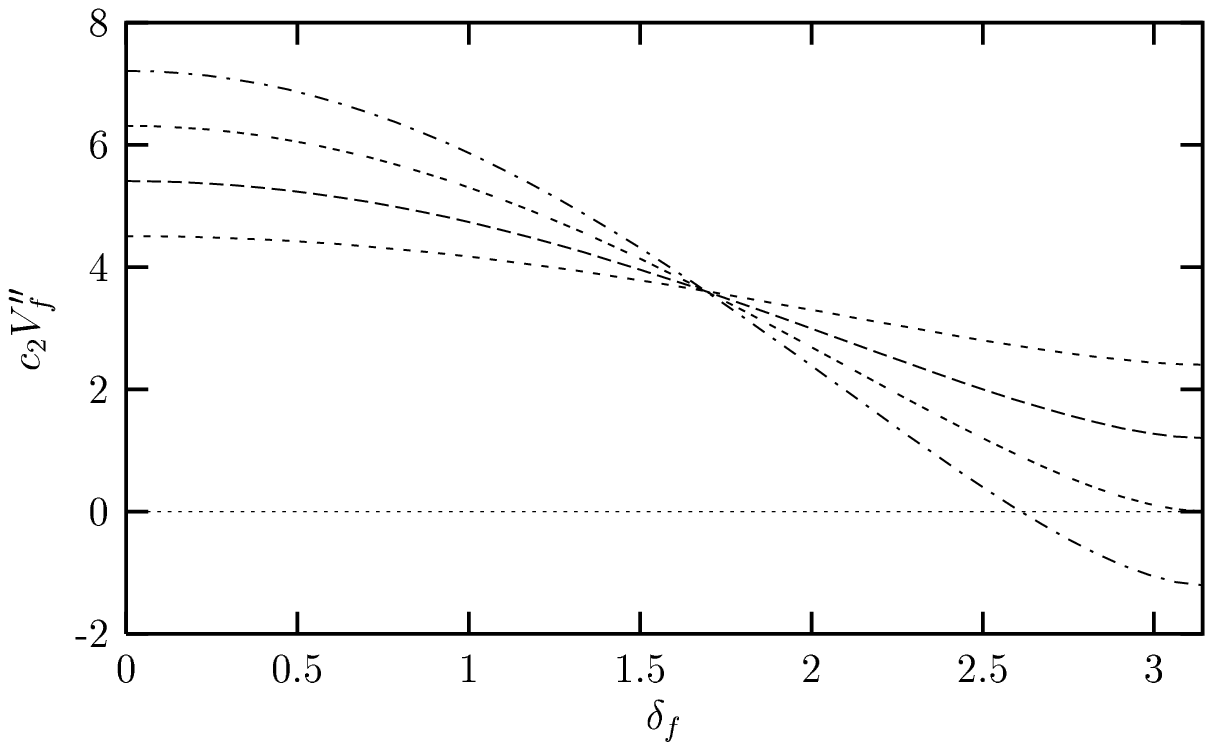}
\vskip -3.15truein 
\caption[]{The $\delta_f$ dependence of $c_2 V_f^{\prime\prime} 
(\theta=\pi, \  N_f, \ z_f=0, \ \delta_f )$, when $N_f=1$ (bottom curve) and  $N_f=4$ (top
curve). The curves in between are  for $N_f=2$, \  $N_f=3$, from
bottom to top.} 
\label{fig8}
\end{figure}

A comparative look at Figure 7 and Figure 8
suggests that    $ c_2 V_f^{\prime\prime} (\theta=0, \, N_f, \ z_f=0, \ \delta_f )$ 
and $c_2 V_f^{\prime\prime} (\theta=\pi, \
N_f, \ z_f=0, \ \delta_f )$ do not change sign with the variation of
$\delta_f$ for $N_f=1,\,\, 2$, whereas they change sign for
$N_f=4$ (the bottom curve  in Figure 7, and the top curve  in Figure 8).

To verify that   $N_f$  indeed   plays a 
role on the structure of the minima, we study the sign of
$V_{f}^{\prime\prime }(\theta, \, N_f, \, z_f=0, \, \delta_f)$ at the limits of $\delta_f$ for each
$\theta=0$, and $\theta=\pi$ as a function of $N_f$ ($\theta=0, \,
\pi$ corresponding to the limits of periodic, and antiperiodic
boundary conditions):
\begin{eqnarray}
c_2 V_{f}^{\prime\prime }(\theta=0, \  N_f, \ z_f=0, \  \delta_f=0) =
\Bigg \{ \bigg ( 3 - N_f \bigg)  \xi_3  \Bigg \} &>& 0 \, \,
\mbox{for}\,\,
N_f=1,2~,\nonumber\\
 &<& 0 \,\,   \mbox{for}\,\,
N_f\geq 4~, \nonumber\\
c_2 V_{f}^{\prime\prime }(\theta=\pi, \ N_f, \ z_f=0, \  \delta_f=\pi) =
\Bigg \{ \bigg ( 3 - N_f \bigg)  \xi_3  \Bigg \} &>& 0 \, \,
\mbox{for}\,\,
N_f=1,2~,\nonumber\\
 &<& 0 \,\,   \mbox{for}\,\,
N_f\geq 4~,
\end{eqnarray}
and
\begin{eqnarray}
c_2 V_{f}^{\prime\prime }(\theta=0, \ N_f, \ z_f=0, \  \delta_f=\pi) =
 \Bigg \{ \frac{3}{4} \xi_3 \bigg ( N_f+4  \bigg)   \Bigg \} &>& 0 \, \, \mbox{for}\,\,
\mbox{all} \,\, N_f~,\nonumber\\
c_2 V_{f}^{\prime\prime }(\theta=\pi, \ N_f, \ z_f=0, \  \delta_f=0) =
\Bigg \{ \frac{3}{4} \xi_3 \bigg ( N_f+4  \bigg)   \Bigg \} &
\simgt & 0 \, \, \mbox{for}\,\, \mbox{all} \,\, N_f.~
\end{eqnarray}

One notices that  for $N_f\geq 4$ case there are some subtleties,
and thus we pay special attention to  $N_f=4$:

We see from Figure 7 that that there is a critical value
$\delta_a^{c_{1}}=0.53$, above which
$c_2 V^{\prime\prime}(\theta=0, \ N_f,  \ z_f=0, \  \delta_f=0 )>0$. 
Moreover as Figure 8 suggests there
is another critical value $\delta_a^{c_{2}}=2.61$ which is
different from the former case, below which
$c_2 V_f^{\prime\prime}(\theta=\pi, \ N_f,  \  z_f=0, \  \delta_f=0)>0$. 
Thus, for  $N_f=4$, the absolute minima are: 
\begin{eqnarray}
&& 0 < \delta_f< \delta_f^{c_{1}} , \,\,\,\theta_m =\pi \nonumber\\
&& \delta_f^{c_{2}}  < \delta_f < \pi ,\,\,\,  \theta_m=0.
\end{eqnarray}
However, as mentioned above for $\theta_m=0,\,\pi$, $U^{sym}=(I,-I)$ and those lie in the center of SU(2), thus the
symmetry is not broken. 

Next, we would like to determine the regions of $\delta_f$, in which
$\theta=0$, $\pi$
are the absolute minima, respectively. 
For this purpose, in Figure 10, and Figure 11, we have plotted  
$c_1 V_{f} (\theta,  \ N_f=1, \   z_f=0, \  \delta_f )$,
with respect to $\theta$, for selected values of $\delta_f$, 
in the $0< \delta_f < \pi/2$, and  $\pi/2 < \delta_f < \pi$
intervals, respectively. 
\begin{figure}[htb]
\vskip -2.6truein 
\centering \epsfxsize=5.5in
\leavevmode\epsffile{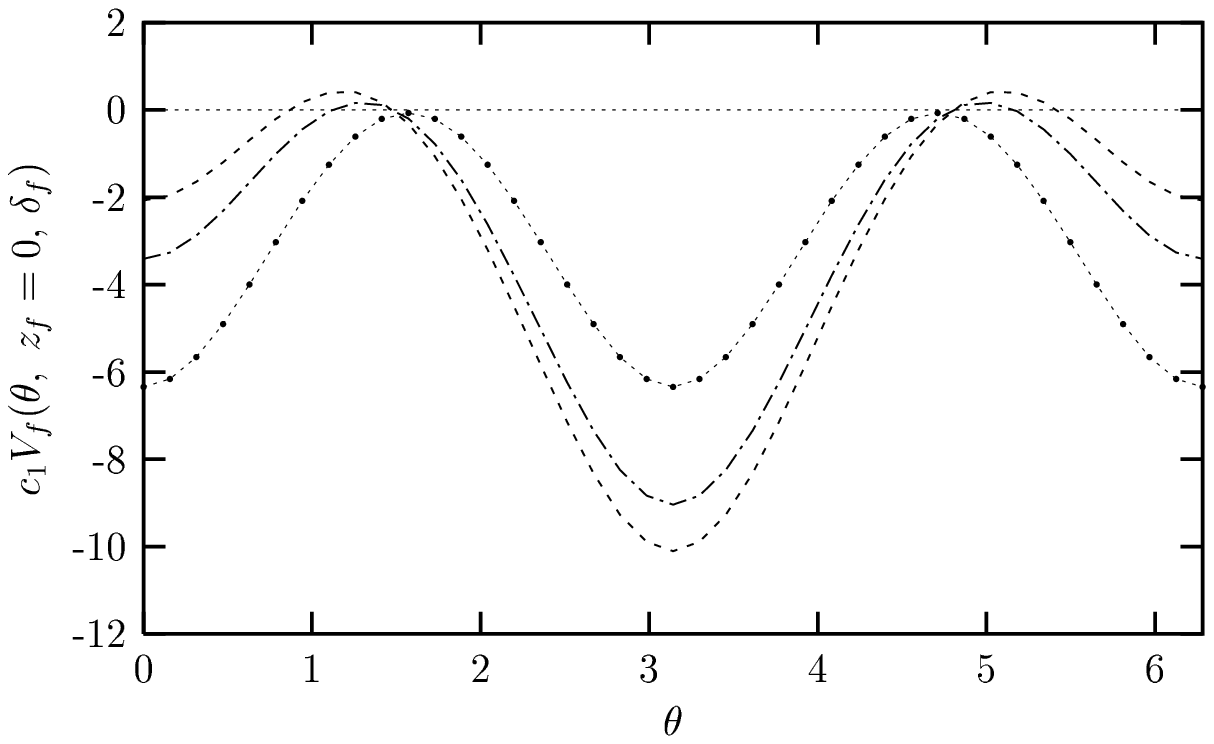} 
\vskip -3.15truein 
\caption[]{The dependence on $\theta$  of
 $c_1 V_{f} (\theta,  \ N_f=1, \   z_f=0,  \, 0< \delta_f < \pi/2)$. 
Here, $\delta_f=0$,  and $\delta_f=  \pi/4$  for the bottom and the middle
curves, respectively, whereas 
$\delta_f=\pi/2$ for the top curve.} 
\label{fig9}
\end{figure}
\begin{figure}[htb]
\vskip -2.7truein 
\centering \epsfxsize=5.5in
\leavevmode\epsffile{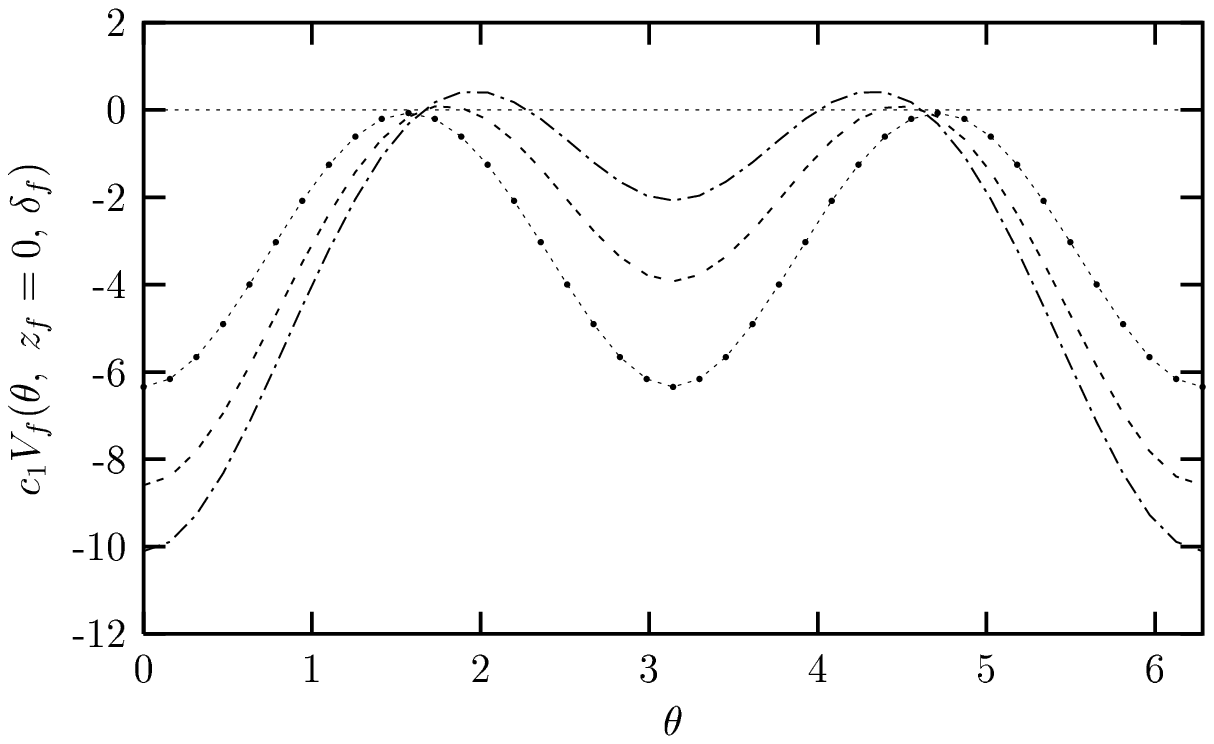}  
\vskip -3.15truein 
\caption[]{The dependence on $\theta$  of  $c_1 V_{f} (\theta,  \ N_f=1, \
z_f=0, \, \pi/2 < \delta_f <\pi)$. Here,    $\delta_f=7 \pi/10$ and   $\delta_f=  \pi$ 
for the middle and top curves,  respectively, whereas $\delta_f=\pi/2$ for
the bottom curve.} 
\label{fig10}
\end{figure}

We see from Figure 9, and 10  that 
$\theta=0$, $\pi$ are the absolute minima for $\pi/2 <\delta_f$, and
$\pi/2< \delta_f< \pi$ respectively, as in the case discussed by
Hosotani~\cite{Hosotani83}, and they are degenerate at $\delta_f=\pi/2$.
However, for $\theta_m=0,\,\pi$, $U^{sym}=(I,-I)$ and these lie in the center of SU(2), thus the
symmetry is not broken.

Similar pattern  can be obtained for the 
behaviour of  $ V_{f} (\theta, \, \  N_f=4, \ z_f=0, \   \delta_f )$
case. We have observed that 
the secondary local minima become shallower (that is smoothed out),
however, with increasing $N_f$.

Next, we have to check the stability properties of the 
absolute minima under the variation of $V_{f}^{\prime\prime }(\theta, \
N_f, \ z_f, \  \delta_f)$ with respect
to $z_f$.
Using  (\ref{fpot}), we get: 
\begin{eqnarray}
c_2 V_{f}^{\prime\prime }(\theta=0, \  N_f, \ z_f, \  \pi/2< \delta_f
<\pi)&=& \sum_{n=1}\frac{1}{n^3} \bigg [3  -\ N_f F (z_f n)
\cos n \delta_{f}  \bigg ]~,\nonumber\\
c_2 V_{f}^{\prime\prime }(\theta=\pi, \  N_f, \ z_f, \  0< \delta_f
<\pi/2)&=& \sum_{n=1}\frac{1}{n^3} \bigg [3  - (-1)^n N_f F(z_f
n) \cos n \delta_{f}   \bigg]~,
\end{eqnarray}
and look for the critical values of $z_f$,  
where $c_2 V_{f}^{\prime\prime } (\theta=0, \  N_f,  \ z_f, \  \pi/2 < \delta_f <\pi)$,
and $c_2 V_{f}^{\prime\prime }(\theta=\pi, \  N_f, \  z_f, \  0< \delta_f
<\pi/2)$ change sign, in Figure  11 and Figure 12, respectively,  
when $N_f=1$. 
\begin{figure}[htb]
\vskip -2.6truein 
\centering \epsfxsize=5.5in
\leavevmode\epsffile{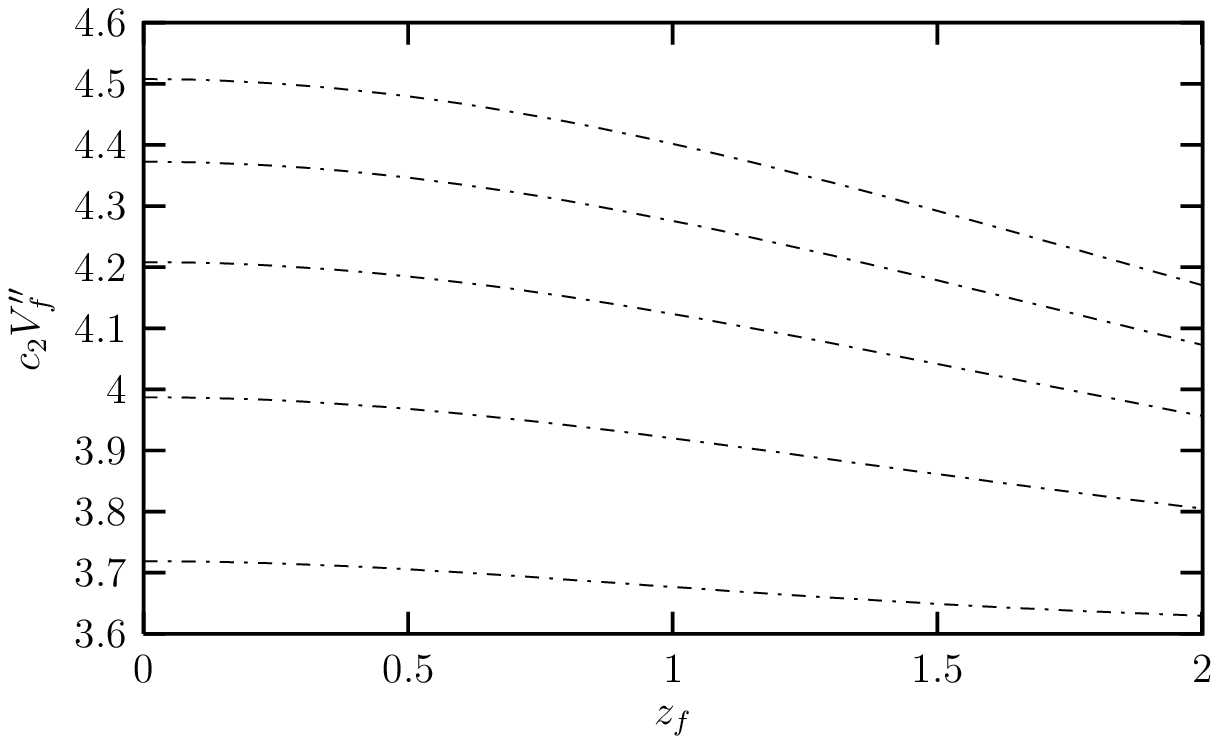} 
\vskip -3.15truein
\caption[]{The dependence on $z_f$  of $c_2 V_{f}^{\prime \prime}
(\theta=0, \ N_f=1, \ z_f, \ \pi/2 < \delta_f < \pi) $.
Here,  $\delta_f=\pi$ for the  top curve,  whereas 
$\delta_f=\pi/2$ for the bottom curve. For the curves in between  
$\delta_f= 6\pi/10, \, 7\pi/10, \, 8\pi/10$,  from bottom to top,
respectively.} 
\label{fig11}
\end{figure}
\begin{figure}[htb]
\vskip -2.6truein 
\centering \epsfxsize=5.5in
\leavevmode\epsffile{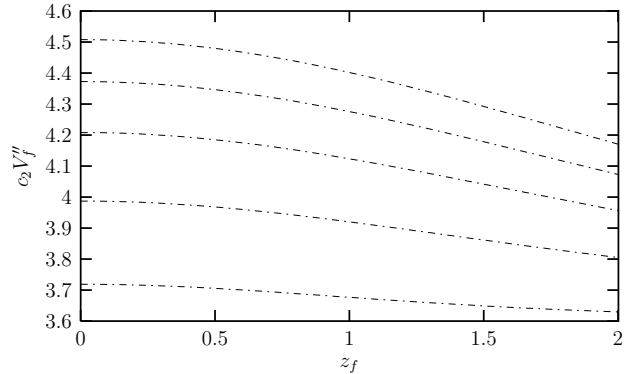}
\vskip -3.15truein
\caption[]{The dependence on $z_f$  of $c_2 V_{f}^{\prime \prime}
(\theta=\pi, \ N_f=1, \ z_f, \ 0 < \delta_f< \pi/2)$. Here,
$\delta_f=0$ for the top curve, whereas 
$\delta_f=\pi/2$, for the bottom curve. For the  curves in between  
$\delta_f= 2 \pi/10, \, 3\pi/10, \, 4\pi/10$, from bottom to top, 
respectively.} 
\label{fig12}
\end{figure}

Note that the plots in Figure 11 and Figure 12 
are identical despite  the fact that 
the intervals for $\delta_f$
are different. This is due to the fact that 
$V_{f}^{\prime\prime }(\theta=0, \  N_f, \  z_f, \  \delta_f )$ $\rightarrow$ 
$V_{f}^{\prime\prime }(\theta=\pi,\  N_f, \ \  z_f, \  \delta_f)$ 
under the transformation $\delta_f \rightarrow \delta_f -\pi$.

A comparative analysis of Figure 11 and Figure 12 shows that 
$V_{f}^{\prime\prime } (\theta, \, N_f=1, \ z_f,\, \delta_f)$ is always positive independent of $z_f$;
that is there are no critical values for  $z_f$. 

Now, we would like to study the behaviour of $V_f(\theta)$ under the
variations of $z_f$. In Figure  13, we have plotted $V_f (\theta,z_f)$
for selected values of $\delta_f$ from the region in which $\theta=0$, and $\pi$
are global minima for $z_f=0$, respectively. 
\begin{figure}[htb]
\vskip -2.7truein 
\centering \epsfxsize=5.6in
\leavevmode\epsffile{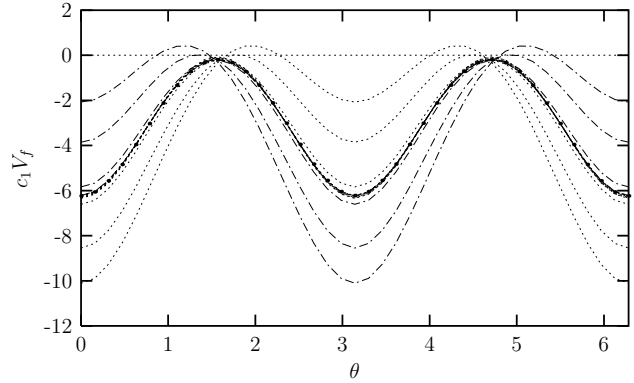} 
\vskip -3.1truein 
\caption[]{The dependence on $\theta$  of $c_1 V_{f} (\theta,  \,  N_f=1, \  z_f, \   \delta_f )$.
Here, $\delta_f=0$,  and   $\delta_f=\pi$ 
are represented by the dot-dashed curves, and dotted curves, respectively.
The dot-dashed curves (with $\delta_f=0$), from  bottom to   top,  are for
$z_f=0$, $z_f= 2$,  and  $z_f=5$, whereas the dotted curves  (with $\delta_f=\pi$), from top to bottom, 
are for $z_f=0$,   $z_f=2$,  and  $z_f=5$.
When  $z_f=8$,  
all the  curves coagulate to the same limiting degenerate 
minima curve (shown by dots in the middle),  which always happens at $\delta_f=\pi/2$.}
\label{fig13}
\end{figure}

We would like to note that, in Figure 13 we have shown the curves for the values of $z_f=0$, $z_f= 2$,  and  $z_f=5$
only, as they are very densly packed and very difficult to distinguish from
each other in the region $5 < z_f < 8$.

We have observed that  for $z_f=8$,
all the curves coagulate to the 
same limiting curve, corresponding to the degenerate minima ($\delta_f=\pi/2$).

Thus, $z_f=8$
emerges as some sort of critical value, not in the sense that we move from 
one global minimum to another, when we cross it; but which ever local minimum we start from, we end up 
with the degenerate minima case, when we reach this value of $z_f$.
As the local minima are always either one of the $\theta=0$, $\pi$,
then there is no change in the symmetry pattern.

Finally, we look at the  variation of $V_f^{\prime\prime} (\theta_m, \ N_f,  \,  z_f, \, \delta_f)$ in the specific intervals of $\delta_f$ found above, for
$N_f=4$: 
For this purpose, in Figure 14 and Figure 15 
we analyze the dependence  of $c_2 V_{f}^{\prime \prime}
(\theta=0, \ N_f=4,   \ z_f, \   \delta_f  )$  on $z_f$  in
the  $\delta_f^{c_{2}}=2.61 < \delta_f < \pi$  and  $0 <\delta_f<
\delta_f^{c_{1}}=0.53$
intervals, respectively, for  selected values of $\delta_f$. 

Note that  Figure 14 and Figure 15 show the same symmetry 
behaviour  we mentioned above,  in relation to the 
Figure 11 and Figure 12.

A comparative analysis of Figure 14 and Figure 15 suggest that  
the variation of $V_f^{\prime\prime}(\theta, \, N_f=4, \,  z_f, \,  \delta_f)$ at $\theta=0$,
$\pi$ with respect to $z_f$,  show similar behaviour to
those of $N_f$=1, 2. 
Thus,  once we restrict $\delta_f$ to the
allowed range in this case, 
mass does not play any further role.
\begin{figure}[htb]
\vskip -2.7truein 
\centering \epsfxsize=5.5in
\leavevmode\epsffile{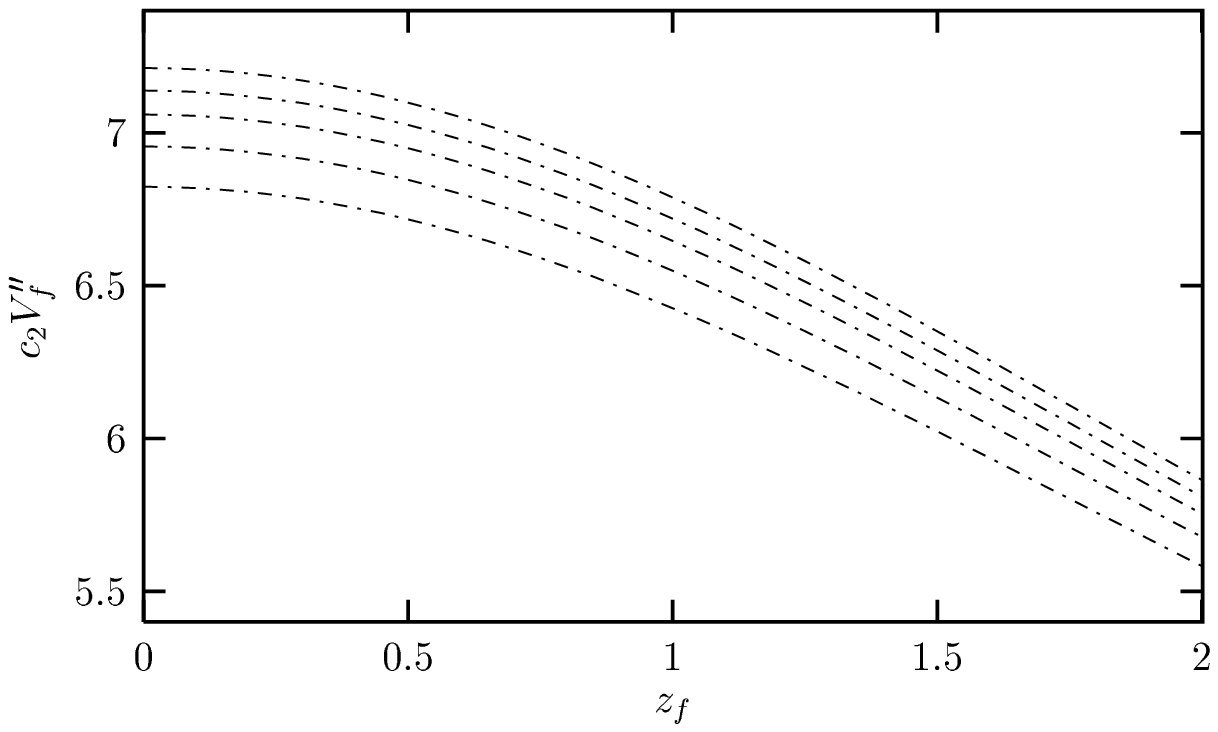} 
\vskip -3.1truein
\caption{The dependence on $z_f$  of $c_2 V_{f}^{\prime \prime}
(\theta=0, \ N_f=4,  \ z_f, \  \delta_f^{c_{2}}  < \delta_f < \pi )$. 
Here, $\delta_f=\pi$ for the top curve,
whereas    $\delta_f=2.61$, for the bottom curve. For the curves in between,
$\delta_f$=2.91, \ 2.81,\  2.71, from bottom to top, respectively.} 
\label{fig14}
\end{figure}
\begin{figure}[htb]
\vskip -2.7truein
\centering \epsfxsize=5.5in
\leavevmode\epsffile{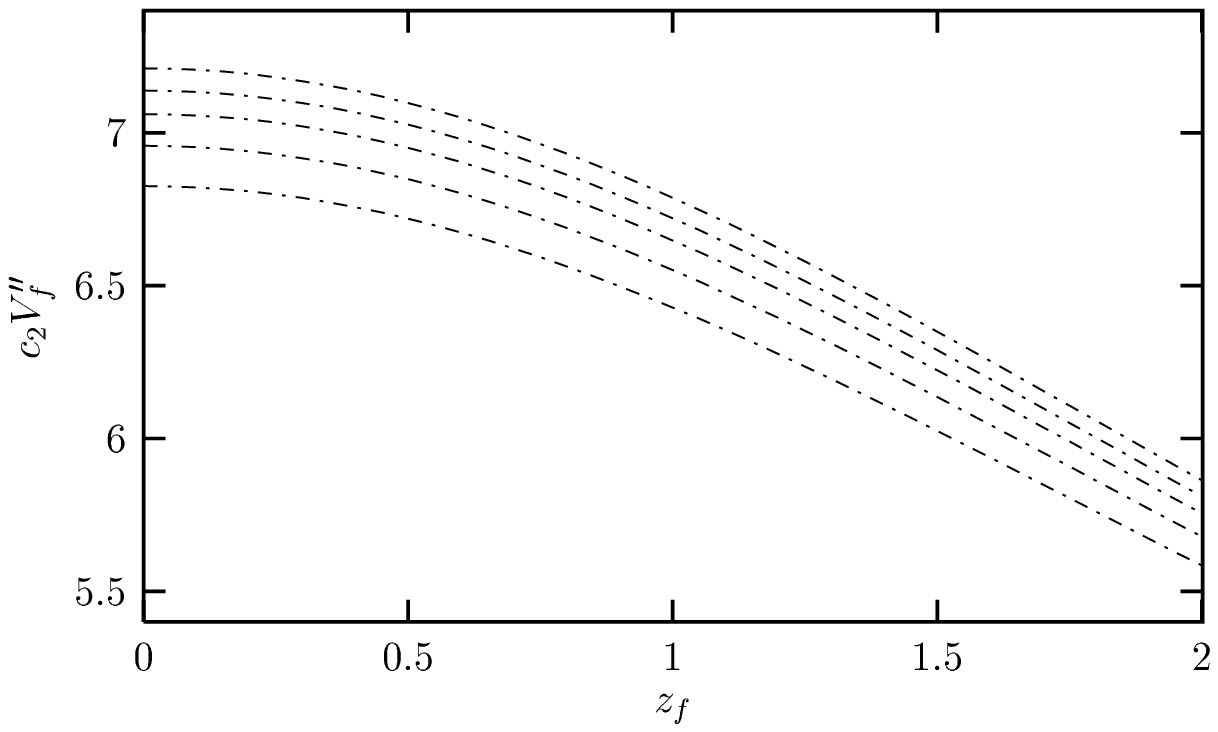} 
\vskip -3.1truein
\caption{The dependence on $z_f$  of $ c_2 V_{f}^{\prime
\prime}(\theta=\pi, \ N_f=4, \ \ z_f, \  0<\delta_f< \delta_f^{c_{1}})$. 
Here,  $\delta_f=0$ for the top curve, whereas
$\delta_f=0.53$, for the bottom curve. For the  curves in between,
$\delta_f$= 0.23, \ 0.33, \ 0.43,  from top to bottom, respectively.} 
\label{fig15}
\end{figure}

In summary the above detailed analysis shows that 
fundamental fermions do not break
SU(2) symmetry, irrespective of the values of the parameters $\delta_f$,
$z_f$, and $N_f$.

\section{Fundamental and Adjoint Fermions with equal masses}

With these inputs in mind, let us look at the general case  where
there are $N_a$ massive adjoint fermions, and $N_f$ massive
fundamental fermions with equal masses ($z_a=z_f=z$). 

The potential is given by:

\begin{eqnarray}
V_{af}(\theta, \ N_a, \ N_f, \ z, \ \delta_a, \  \delta_f)&=&
\frac{1}{c_1} \sum_{n=1}  \frac{1}{n^5}  \Bigg \{ \bigg[-3+ 4 \
N_a F (zn)  \cos n
\delta_a \bigg]  \bigg(1+  \cos 2  n \theta \bigg) \nonumber\\
&+& \ 4 \ N_f  F (zn) \cos n \delta_{f}  \cos n  \theta \Bigg \}~.
\label{afpot}
\end{eqnarray}
The most trivial roots of $V^{\prime}_{af}$ are $\theta=0, \, \pi$. In principle,  there could 
be non-trivial roots of $V^{\prime}_{af}=0$ as well,  depending on the
values of the parameters; we checked this numerically, 
and analytically, and have shown that 
there are no other minima. Again, as in the previous special cases
we look at the two special limits, namely $z_a=z_f \rightarrow
\infty$, and $z_a=z_f\rightarrow 0$.

One first notes that for $m_a=m_f \rightarrow \infty$, \,\, \,  
$F (zn)  \rightarrow 0$ which means  $V \rightarrow V_{\rm {pure\
gauge}}$, as
all fermions, adjoint and fundamental, decouple (thus, as before  
$\theta=0 \, [\mbox{mod} \, \, \pi]$ is an  absolute minimum).
For   $m_a=m_f \rightarrow 0$, $F (zn) \rightarrow 1$, and 
\begin{eqnarray}
c_2 V_{af}^{\prime\prime} (\theta=0, \ N_a, \ N_f, \ z=0, \,
\delta_a, \, \delta_f)&=&
 \sum_{n=1}  \frac{1}{n^3} \bigg(3- 4 \ N_a   \cos n \delta_a -  N_f   \cos
n \delta_{f} \bigg)~,\nonumber\\
c_2 V_{af}^{\prime\prime} (\theta=\pi, \ N_a, \ N_f, \  z=0, \, \delta_a,
\ \delta_f)&=&
 \sum_{n=1}  \frac{1}{n^3} \bigg(3- 4 \ N_a   \cos n
\delta_a -  N_f (-1)^n  \cos n \delta_{f} \bigg)~.
\end{eqnarray}
To determine the ranges of $\delta_{f}$, $\delta_{a}$, 
we first plot  the 
$\delta_f-\delta_a$ region, for which $c_2 V_{af}^{\prime \prime}
(\theta=0, \ N_a, \ N_f,\  z=0,  \  \delta_a, \ \delta_f)>0$ in Figure 16,
when $N_a=N_f=1$.
\begin{figure}[htb]
\vskip -2.7truein 
\centering \epsfxsize=5.5in
\leavevmode\epsffile{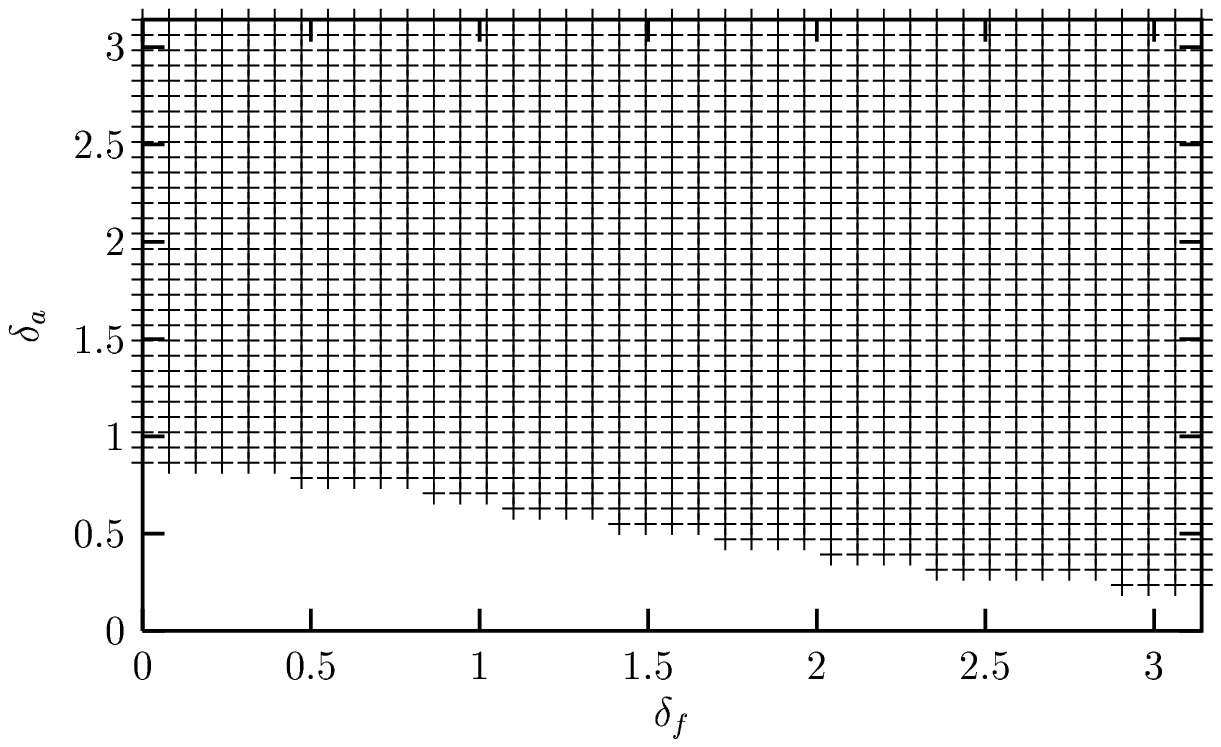} 
\vskip -3.1truein 
\caption{The  $\delta_f-\delta_a$
region, for which $c_2 V_{af}^{\prime
\prime}(\theta=0, \ N_a=N_f=1, \ z=0, \  \delta_a, \ \delta_f)>0$.} 
\label{fig16}
\end{figure}

As can be observed from  Figure 16 that the lower bound of $\delta_a$ ranges
from $\delta_a= 3 \pi/40$, up to  $\delta_a= \pi/4$,
when $\delta_f$ changes from 0 to $\pi$.
One notes that the lower bound of $\delta_f$ ranges from  $ 3 \pi/20$ to $37
\pi/40$, in the  $3 \pi/40 \simlt \delta_a  \simlt  \pi/4$,
interval.
On the other hand, for  $10 \pi/ 40 <\delta_a < \pi$, 
there is no constraint on $\delta_f$; 
That is, all values of $\delta_f$ are allowed for $\delta_a> 10\pi/40$.
\begin{figure}[htb]
\vskip -2.7truein 
\centering \epsfxsize=5.5in
\leavevmode\epsffile{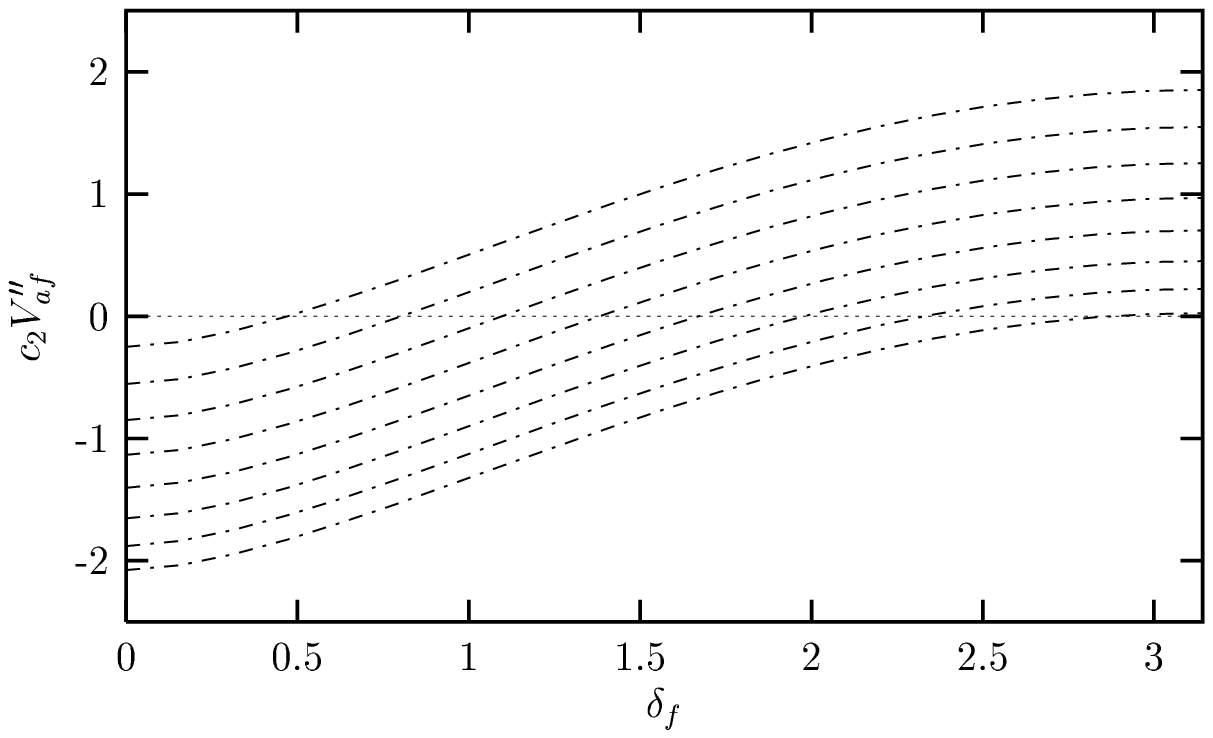} 
\vskip -3.1truein 
\caption{The dependence on $\delta_f$  of $c_2 V_{af}^{\prime\prime}$ $(\theta=0,
\  N_a=N_f=1, \   z=0, \, \delta_a, \,  \, \delta_f)$,  for
$3 \pi/40 \simlt \delta_a \simlt \pi/4$. Here, $\delta_a=\pi/4$ for the top
curve, whereas $\delta_a=3 \pi/40$, for  the bottom curve. 
In the remaining portion of the parameter
space, namely  $\delta_a < 3 \pi/40$, and  $\delta_a >
\pi/4$,  $c_2 V_{af}^{\prime\prime}$ does not change sign.}
\label{fig17}
\end{figure}

In Figure 17 we plot  $c_2 V_{af}^{\prime\prime} (\theta=0,\  N_a=N_f=1, \ z=0, \, \delta_a,
\, \delta_f)$  with respect to  $\delta_f$, for the set of values $ 3 \pi/40\simlt \delta_a \simlt \pi/4$,
for which $c_2 V_{af}^{\prime\prime} (\theta=0, \ N_a= N_f=1, \ z=0, \, \delta_a,
\, \delta_f)$ was changing sign in Figure 16. 
Here,  $\delta_a=\pi/4$ for the top curve, whereas  $\delta_a=3 \pi/40$ for the bottom curve. 

As can be seen from Figure 17 that  when  $\delta_a=\pi/4$,
$c_2 V_{af}^{\prime\prime} (\theta=0, \ N_a=N_f=1,\ z=0, \, \delta_a,
\, \delta_f)$ changes sign at  $\delta_f= 3  \pi/20$,
whereas   for  $\delta_a=3 \pi/40$, the  lower bound  on  $\delta_f$ moves to  $\delta_f=37 \pi/40$. 
For $\delta_a > \pi/4$, $c_2
V_{af}^{\prime\prime} (\theta=0, \ N_a=N_f=1, \ z=0, \, \delta_a,
\, \delta_f)$ does not change sign for any value of $\delta_f$,
which is consistent with Figure 16. 
\begin{figure}[htb]
\vskip -2.6truein 
\centering \epsfxsize=5.5in
\leavevmode\epsffile{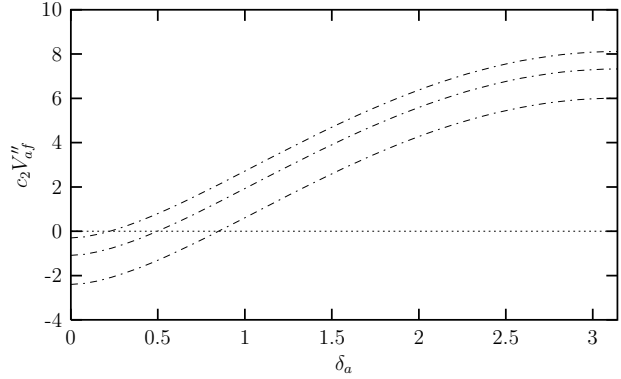} 
\vskip -3.1truein 
\caption{The dependence on $\delta_a$  of
$c_2 V_{af}^{\prime\prime} (\theta=0, \ N_a= N_f=1, \ z=0, \, \delta_a, \, \delta_f)$.
Here,  $\delta_f=\pi$,
$\delta_f=\pi/2$ and $\delta_f=0$, for the top, middle, bottom curves, respectively. 
$c_2 V_{af}^{\prime \prime}$ changes sign for all values of $\delta_f$;
the critical values of $\delta_a$ decrease with increasing values of
$\delta_f$.}
\label{fig18}
\end{figure}

In Figure 18,  we analyze the dependence on $\delta_a$  of
$c_2 V_{af}^{\prime\prime} (\theta=0, \  N_a= N_f=1, \  z=0, \, \delta_a, \, \delta_f)$ 
for given values of $\delta_f$, namely  $\delta_f=\pi$ (top curve),
$\delta_f=\pi/2$ (middle curve), and $\delta_f=0$ (bottom curve). 

As can be seen  from Figure 18,   
$c_2 V_{af}^{\prime \prime}(\theta=0)$ changes sign for all values of $\delta_f$,
whereas the lower bound on $\delta_a$ moves from $ \pi/4$ to $ 3 \pi/40$,
with the increasing values of $\delta_f$. 
For instance, when $\delta_f=0$  the  lower bound on  $\delta_a$,
at which the $c_2 V_{af}^{\prime\prime}$ changes sign,
is $\delta_a=\pi/4$, 
whereas  for   $\delta_f=\pi$, it is  $\delta_a=3 \pi/40$,
which is consistent with Figure 16.
\begin{figure}[htb]
\vskip -2.6truein 
\centering \epsfxsize=5.5in
\leavevmode\epsffile{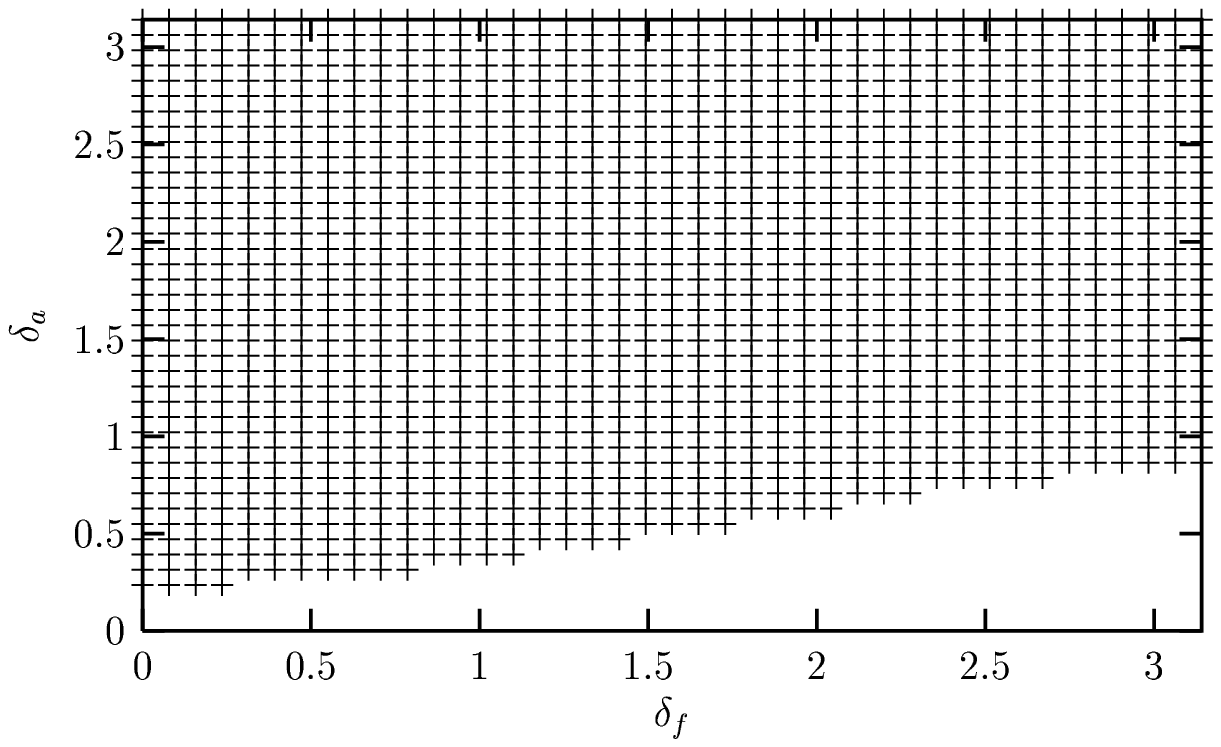} 
\vskip -3.1truein 
\caption{The  $\delta_f$-$\delta_a$ region,  for which $V_{af}^{\prime
\prime}(\theta=\pi, \ N_a=N_f=1, \ z=0, \  \delta_a, \ \delta_f ) >0$.} 
\label{fig19}
\end{figure}

In Figure 19,  we plot the 
$\delta_f$-$\delta_a$ region, for
which  $c_2 V_{af}^{\prime\prime} (\theta=\pi,  \ N_a=N_f=1, \ z=0,
 \  \delta_a, \ \delta_f )>0$. Here,
all values of $\delta_f$ from 0 to $\pi$ are allowed above the
critical value  $\delta_a> \pi/4$, as was the case in Figure 16, also.

In Figure 20, we plot  $c_2 V_{af}^{\prime\prime} (\theta=\pi, \ N_a=N_f=1, \ z=0,
 \, \delta_a, \, \delta_f)$ with respect to
$\delta_f$,
for the set of values $ 3 \pi/40\simlt \delta_a \simlt \pi/4$,
in the region where  $c_2 V_{af}^{\prime\prime} (\theta=\pi, \ N_a=N_f=1,\
z=0,  \, \delta_a,
\, \delta_f)$ was changing sign in Figure 19. 
In Figure 20, the top  and the bottom curves
represent $\delta_a=\pi/4$, and 
$\delta_a=3 \pi/40$, respectively. 
One notes that $c_2 V_{af}^{\prime\prime} (\theta=\pi,  \ N_a=N_f=1, \ z=0,  \, \delta_a,
\, \delta_f)$ does not change sign
in the remaining portion of the $\delta_a$-parameter
space (namely,  for $\delta_a <3 \pi/40$, and  $\delta_a>\pi/4$). 
\begin{figure}[htb]
\vskip -2.6truein 
\centering \epsfxsize=5.5in
\leavevmode\epsffile{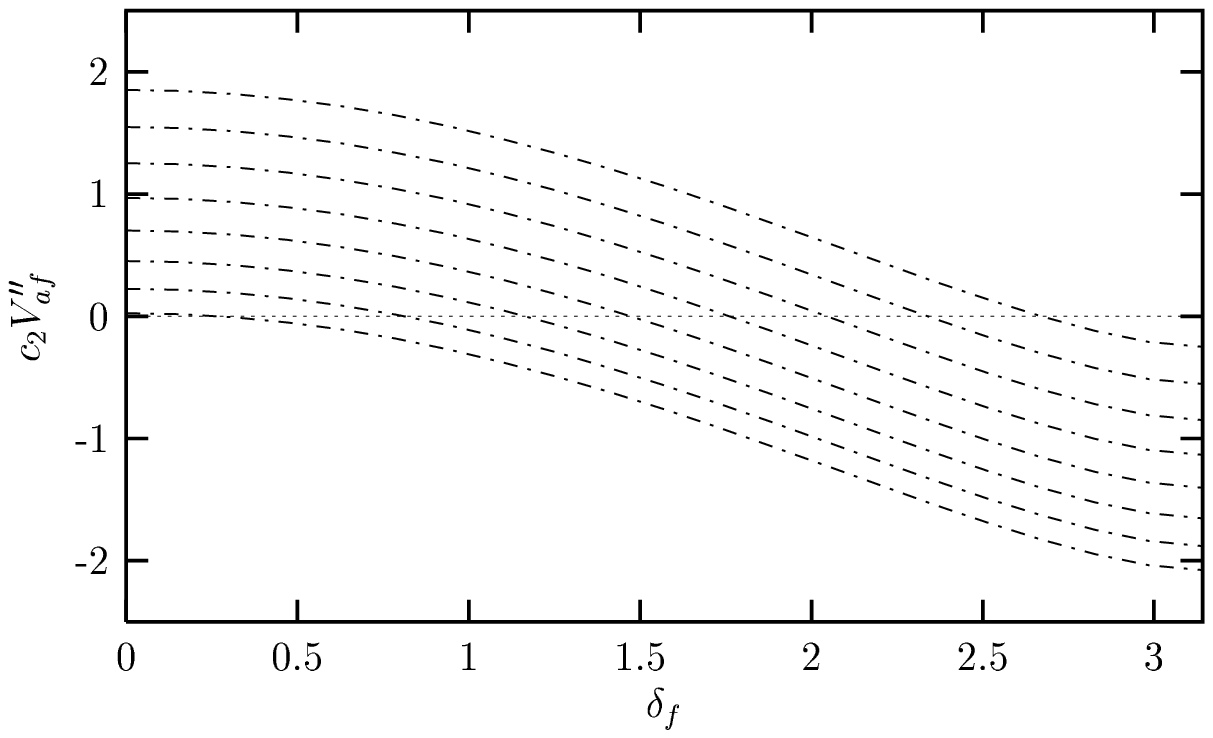} 
\vskip -3.1truein
\caption{The dependence on $\delta_f$  of $c_2
V_{af}^{\prime\prime} (\theta=\pi,  \ N_a=N_f=1, \ z=0,  \, \delta_a,
\, \delta_f)$,  for the set of values $3 \pi/40\simlt \delta_a
\simlt \pi/4$. Here,  $\delta_a=\pi/4$ for the
top curve, whereas  $\delta_a=3 \pi/40$, for the bottom curve. In the remaining
portion of the $\delta_a$-parameter space  $c_2 V_{af}^{\prime\prime}$ does not
change sign.} 
\label{fig20}
\end{figure}

In Figure 21,  we analyze the dependence on $\delta_a$  of
$c_2 V_{af}^{\prime\prime} (\theta=\pi, \ N_a=N_f=1, \ z=0, \, \delta_a, \, \delta_f)$ 
for given values of $\delta_f$ when $N_f=N_a=1$.
In the Figure,  the top, middle,  bottom  curves represent $\delta_f=0$,
$\delta_f=\pi/2$ and   $\delta_f=\pi$, respectively.  
Similar to observations made for Figure 20,
when $\delta_f=\pi$, the potential changes sign at $\delta_f= \pi/4$, 
whereas for $\delta_f=0$, the lower bound on $\delta_a$   
is  $\delta_f= 3 \pi/40$, which is consistent with Figure 19.
\begin{figure} [htb]
\vskip -2.6truein 
\centering \epsfxsize=5.5in
\leavevmode\epsffile{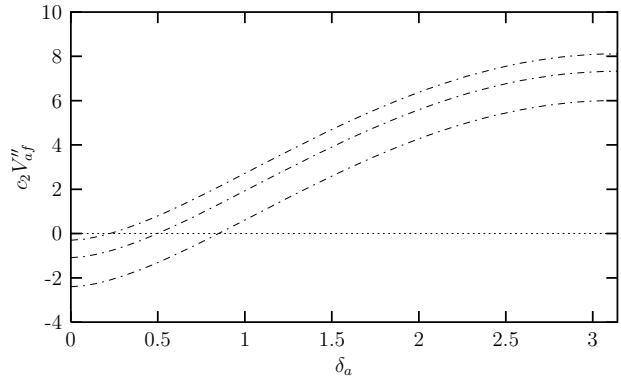} 
\vskip -3.1truein
\caption{The dependence on $\delta_a$  of
$c_2 V_{af}^{\prime\prime} (\theta=0,  \ N_a=N_f=1, \ z=0,  \, \delta_a, \, \delta_f)$.
Here, $\delta_f=\pi$,
$\delta_f=\pi/2$ and   $\delta_f=0$,  for the top, middle and the bottom
curves, respectively.}
\label{fig21}
\end{figure}

We have to next check the region of $\delta_a-\delta_f$
for which $\theta=0$, $\pi$
are the absolute minima. 
For this purpose,
in  Figure 22  we have plotted
$c_2 V_{af} (\theta,  \ N_a=N_f=1, \ z=0,  \, \delta_a, \, \delta_f )$,
with respect to $\theta$ for selected values of $\delta_a >
\delta_a^{cr}=\pi/4 $, identified in Figure 16 and Figure 19
for both  $\theta=0$, and $\pi$.
\begin{figure}[htb]
\vskip -2.8truein 
\centering \epsfxsize=5.6in
\leavevmode\epsffile{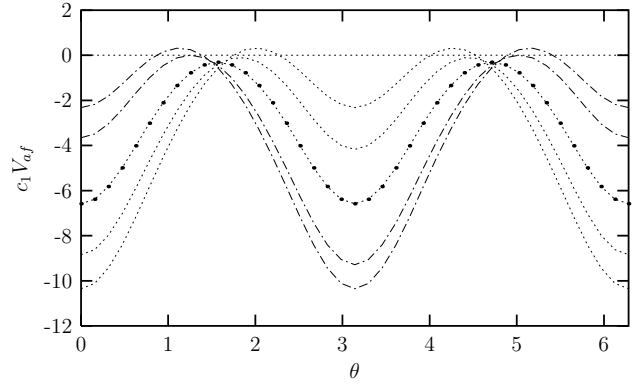} 
\vskip -3.1truein
\caption{The dependence  of $c_1 V_{af} (\theta,  \ N_a=N_f=1, \
z=0,  \, \delta_a=\pi/2, \,  \delta_f)$ on  $\theta$. 
Here, the dotted-line in the middle is for $\delta_f=\pi/2$.
The upper two lines, from top to bottom, are for 
$\delta_f=\pi$, and $\delta_f= 7 \pi/10$, respectively,
chosen in the  $\pi/2< \delta_f< \pi$ interval. 
The lower two lines are for, from bottom to top,  $\delta_f=0$, and
$\delta_f= \pi/4$, respectively, chosen in the  $0< \delta_f <\pi/2$
interval.}
\label{fig22}
\end{figure}

It is seen from Figure 22  that  
 $\theta=0$ is the absolute minimum for $\pi/2 <\delta_f <\pi$,
whereas $\theta=\pi$ is the absolute minimum for  $0 <\delta_f <\pi/2$,
as in the pure fundamental fermions case. 
Note that these are degenerate at  $\delta_f =\pi/2$.
However, as before irregardless of which one of  $\theta=0$, $\pi$
are the absolute minima, the symmetry is unbroken, provided that 
$\delta_a>\delta_a^{cr}=\pi/4$.

Similar analysis  can be made for 
 $c_1 V_{af} (\theta,  \ N_a=N_f=1, \
z=0,  \, \delta_a, \, \delta_f=\pi/2)$,
for selected  values of $\delta_a$,
chosen from the region $\delta_a>\delta_a^{cr}=\pi/4$.
We see that  in this case $\theta=0$, $\pi$
are the degenerate absolute minima,  which is consistent with the remarks of Figure
16 and Figure 19.

Next, we would like to adress the issue of stability 
of these absolute minima,  we have found for the
massless case,  under the variations of z.

We have previously observed 
that there were critical values of z, at which symmetry pattern changed,
for the adjoint case.
But in the fundamental fermions case the mass did not play any role on the
symmetry pattern.
In the present case, that is when the fundamental and adjoint fermions
exist together (with equal masses),
we would like to check which behaviour of the Äprevious special cases would
be carried over.

We get from Eq. (\ref{afpot}):

\begin{eqnarray}
c_2 V_{af}^{\prime\prime} (\theta=0, N_a, \ N_f,\  z, \, \delta_a, \,
\delta_f)&=&  \sum_{n=1}  \frac{1}{n^3} \bigg(3- 4 \ N_a F (z n)
\cos n
\delta_a  \nonumber\\
&-& \  \ N_f  F (z n) \cos n \delta_{f}\bigg)~,\nonumber
\end{eqnarray}
\begin{eqnarray}
c_2 V_{af}^{\prime\prime} (\theta=\pi, \ N_a, \ N_f, \ z, \, \delta_a, \,
\delta_f)&=&
 \sum_{n=1}  \frac{1}{n^3} \bigg(3- 4 \ N_a F (z n)  \cos n
\delta_a    \nonumber\\
&-& \  \ N_f (-1)^n  F(z n) \cos n
\delta_{f}\bigg)~.
\end{eqnarray}

In Figure 23, and Figure 24,  we plot the 3-dimensional graphs depicting  $c_2 V_{af}^{\prime
\prime} (\theta=0, \ N_a=N_f=1,$ 
$ \  z, \, \delta_a, \, \delta_f)$, 
against  $\delta_a-z$, and  $\delta_f-z$,  
when  $\delta_a-\delta_f$ are 
restricted to the region where $\theta=0$
was a minimum in the massless case.

In Figure 23, we obtain 3-dimensional surfaces for each value of $\delta_f$,
which we choose within the  interval [0,$\pi$].
For instance the top surface corresponds to $\delta_f=\pi$,
whereas the bottom one represents  $\delta_f=0$. One notes that 
for   $\delta_f=0$, the lowest allowed bound of $\delta_a$  
is $\delta_a= 11 \pi/40$, consistent with  Figure 16.
\begin{figure}[htb]
\vskip -2.6truein
\centering
\epsfxsize=6.4in
\epsffile{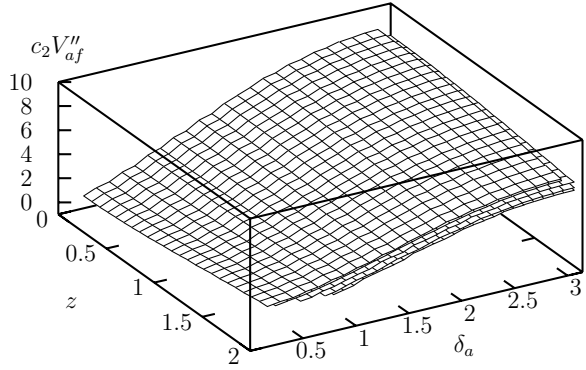}
\vskip -3.6truein
\caption{The dependence of  $c_2 V_{af}^{\prime
\prime} (\theta=0, \ N_a= N_f=1, \  z, \, \delta_a, \, \delta_f)$ 
on $\delta_a$  and z, for selected values of $\delta_f$. Here,
$\delta_a$ and $\delta_f$ 
values are restricted to the shaded region in Figure 16.}
\label{fig23}
\end{figure}
\vskip 3.0cm
\begin{figure}[htb]
\vskip -3.1truein
\centering
\epsfxsize=6.4in
\epsffile{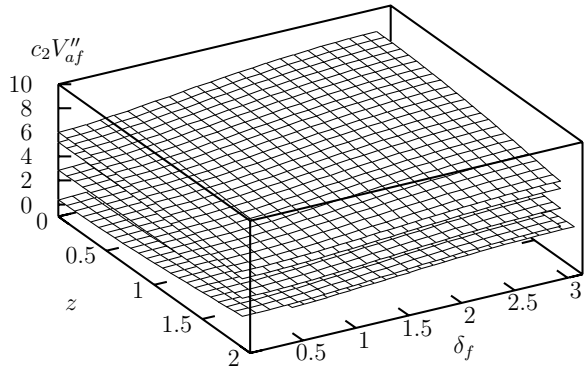}
\vskip -3.6truein
\caption{The dependence of  $c_2 V_{af}^{\prime
\prime} (\theta=0, \ N_a=N_f=1, \  z, \  \delta_a, \, \delta_f)$ 
on $\delta_f$  and z, for selected values of $\delta_a$. Here,
$\delta_a$ and $\delta_f$ 
values are restricted to the shaded region in Figure 16.}
\label{fig24}
\end{figure}

Similarly, each three dimensional surface in Figure 24
corresponds to a discrete value of $\delta_a$
changing in the [0,$\pi$] interval. Here,
the bottom surface represents  $\delta_a=\pi/4$,
at which case the lowest allowed bound on $\delta_f$ is $\delta_f=3 \pi/20$,
as  in Figure 16.

A comparative analysis of Figure 23 and Figure 24 shows that  
$c_2 V_{af}^{\prime
\prime} (\theta=0, \ N_a=N_f=1,\  z, \, \delta_a, \, \delta_f)$, 
 does not change sign with the variations of z,
when $\delta_a$, and $\delta_f$
are restricted to the shaded region in Figure 16.
That is $\theta=0$
remains as a minimum independent of the 
values of z.

To address the  stability issue of the  minimum $\theta=\pi$,
we  plotted a similar set of 3-dimensional graphs 
depicting  $ c_2 V_{af}^{\prime \prime} (\theta=\pi, \
\ N_a=N_f=1, \  z,  \, \delta_a, \, \delta_f)$, 
against $\delta_a-z$ in Figure 25, and 
$\delta_f-z$ in Figure 26, where $\delta_a$ and $\delta_f$ 
values are restricted to the shaded  region in Figure 19.
\begin{figure}[htb]
\vskip -2.6truein
\centering
\epsfxsize=6.4in
\epsffile{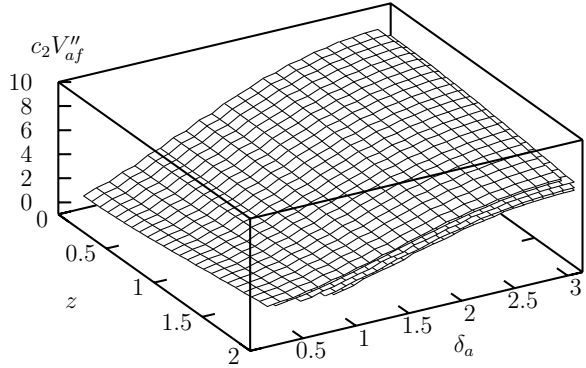}
\vskip -3.6truein
\caption{The dependence of  $c_2 V_{af}^{\prime
\prime} (\theta=\pi, \ N_a=N_f=1, \  z,  \, \delta_a, \, \delta_f)$, 
on $\delta_a$  and z, for selected values of $\delta_f$. Here,
$\delta_a$ and $\delta_f$ 
values are restricted to the shaded region in Figure 19.}
\label{fig25}
\end{figure}
\vskip 3cm
\begin{figure}[htb]
\vskip -3.1truein
\centering
\epsfxsize=6.4in
\epsffile{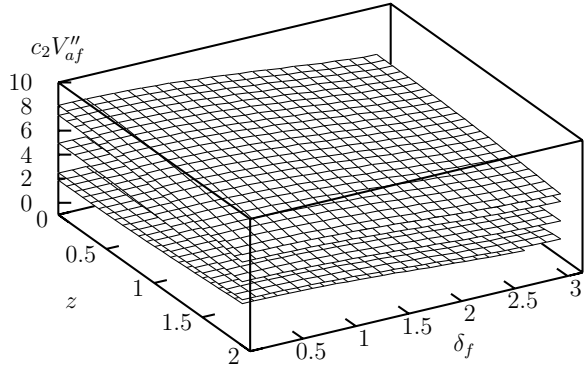}
\vskip -3.6truein
\caption{The dependence of  $c_2 V_{af}^{\prime
\prime} (\theta=\pi, \ N_a= N_f=1, \  z, \, \delta_a, \, \delta_f)$, 
on $\delta_f$  and z, for selected values of $\delta_a$.
Here,
$\delta_a$ and $\delta_f$ 
values are restricted to the shaded region in Figure 19.}
\label{fig26}
\end{figure}

In Figure 25,  the top surface corresponds to $\delta_f=0$,
and the bottom one represents  $\delta_f=\pi$, whereas 
in Figure 26,  the top surface corresponds to $\delta_a=\pi$,
and  the bottom one to  $\delta_a=\pi/4$. 

A comparative analysis of Figure 25 and Figure 26 shows that  $ c_2
V_{af}^{\prime \prime} (\theta=\pi, \ N_a=N_f=1, \  z, \  \delta_a, \, \delta_f)$
does not change sign, and always stays positive,
independent of the values of z, in the allowed region  of $\delta_a-\delta_f$.

Now, we would like to investigate the special case in which $\theta=0$
and $\theta=\pi$
minima are degenerate which can be checked easily to occur  for $\delta_f=\pi/2$,
and $\delta_a>\delta_a^{cr}$.
The result is depicted in Figure 27, where 
we plot $ c_2 V_{af}^{\prime \prime} 
(\theta=0 \ [\mbox{mod} \pi], \ N_a=N_f=1, \  z,  \, \delta_a, \, \delta_f)$
with respect to z
for $\delta_f=\pi/2$, and for the selected set of values of $\delta_a$ within the
allowed range;  $\delta_a> \delta_a^{cr}=\pi/4$.
Namely, we choose   $\delta_a=\pi$ (top curve), $\delta_a= 3  \pi /2$ (middle
curve), $\delta_a= \pi/2$  (bottom curve).
\begin{figure}[htb]
\vskip -2.6truein
\centering
\epsfxsize=5.5in
\epsffile{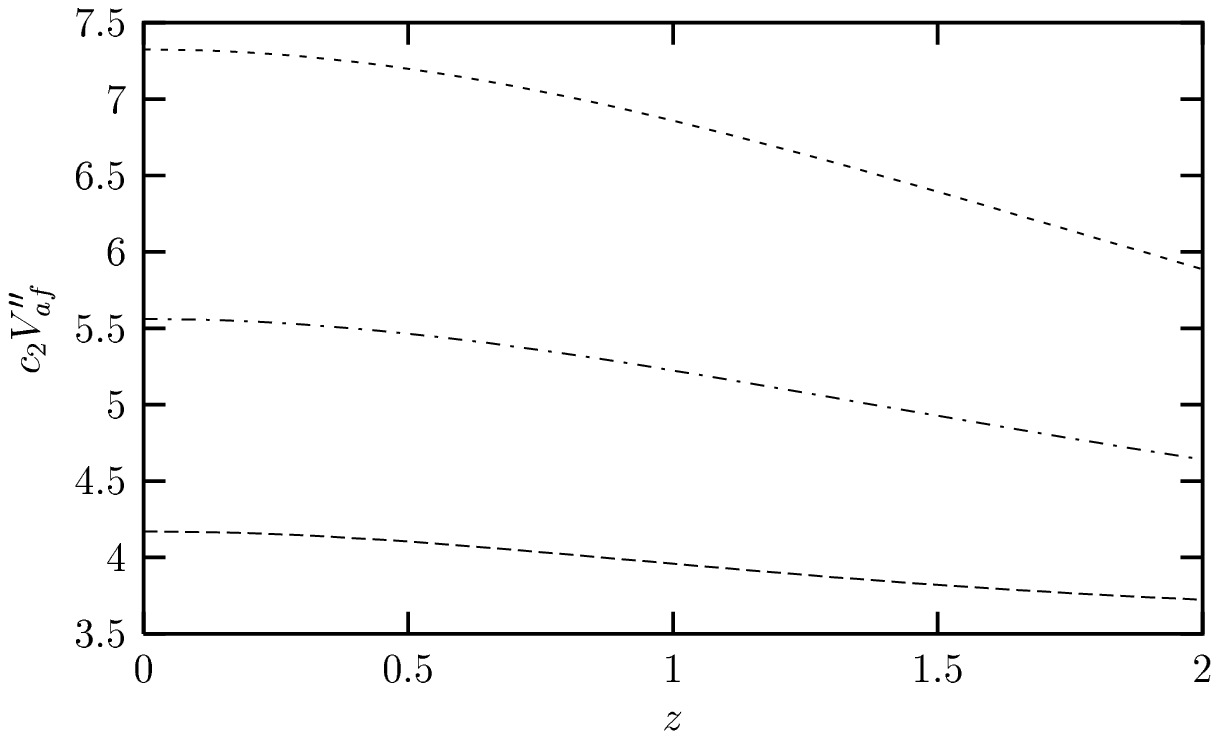}
\vskip -3.1truein
\caption{The dependence on z  of $ c_2 V_{af}^{\prime \prime} 
(\theta=0 \ [\mbox{mod} \pi], \ N_a=N_f=1, \  z,  \, \delta_a, \, \delta_f)$  when 
$\delta_f=\pi/2$, for the selected set of values of $\delta_a$:
$\delta_a=\pi$ (top curve), $\delta_a=3  \pi/2$ (middle curve), 
$\delta_a= \pi/2$  (bottom curve).} 
\label{fig27}
\end{figure}

We see that for this special regime of the parameters, namely $\delta_f=\pi/2$,
 $\delta_a > \delta_a^{cr}$,   $V_{af}^{\prime \prime}(\theta=0,\pi)$ 
is always positive (does not change sign)
with the variations of z.

Lastly, we would like to study the behaviour of the effective potential 
under the variations of z. 
For this purpose, in Figure 28, we have plotted $V_{af}$ ($\theta$,z)
for selected characteristic values of $\delta_a$
and $\delta_f$
picked from Figure 22 (which in turn identified from Figure 16 and 19),
defining the allowed regions of the parameter space for the local minima
$\theta_m=0,\pi$.
\begin{figure}[htb]
\vskip -2.6truein
\centering
\epsfxsize=5.6in
\leavevmode\epsffile{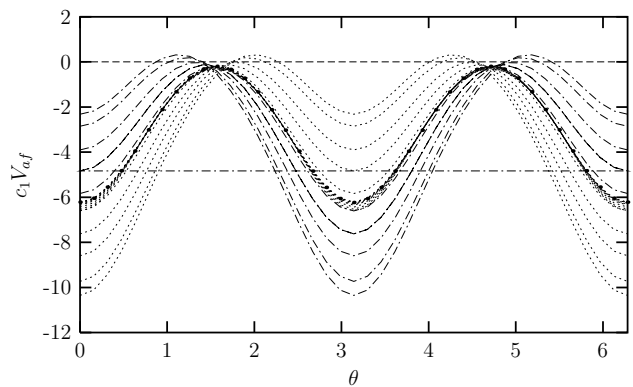}
\vskip -3.1truein
\caption[]{The dependence on $\theta$  of
$V_{af}$ when $N_f=1$, $N_a=1$, for the cases 
$\delta_a=\pi/2$, $\delta_f=\pi/2$ (shaded region),
$\delta_a=\pi/2$, $\delta_f=0$ (dot-dashed curves),
$\delta_a=\pi/2$, $\delta_f=\pi$ (dotted curves).
Here, the dot-dashed lines, from  bottom to   top,  are for
$z=0$, $z= 1$,  $z=2$, $z=3$, $z=5$.
The dotted lines, from top to bottom, are for $z=0$, $z= 1$,  $z=2$, $z=3$, $z=5$.
Again when  $z_f=8$, all the  curves coagulate to the same limiting degenerate minima curve
(shown by dots on the edge of the shaded region), 
which always happens at $\delta_f=\pi/2$, irregardless of the values of $\delta_a$.}
\label{fig28}
\end{figure}

We see from Figure 28 that, for the pair $\delta_a=\pi/2$, $\delta_f=\pi$,
we get the dotted curves  for the values of  $z=0$, $z= 1$,  $z=2$, $z=3$,
$z=5$, from top to bottom,  with  the global minimum $\theta_m=0$.
Similarly, for $\delta_a=\pi/2$, $\delta_f=0$,
we obtain the dot-dashed curves  for the values of  $z=0$, $z= 1$,  $z=2$, $z=3$,
$z=5$, from bottom to top,  with  the global minimum $\theta_m=\pi$.

We see that these two groups of curves (with   $\theta_m=0$, and
$\theta_m=\pi$, respectively)
all coagulate to the limiting curve shown by dots on the edge of the shaded
region, which corresponds to the degenerate minima case (for $\delta_f=\pi/2$
irregardless of the values of  $\delta_a$).

Thus $z=8$ emerges as some critical value, not in the sense that we move from one global
minimum to another, when we cross this value;  but whichever local minimum we start from (in the
massless case) we end up with the degenerate minima case when we reach
this value of z, as in the pure fundamental fermions  case. 
As the absolute minima are always either one of the $\theta=0,\pi$,
then there is no change in symmetry pattern as a consequence of this rather
intricate dynamical phenomenon.

The above detailed analysis shows that  the absolute minima of the massless case,
namely $\theta=0$ or $\theta=\pi$
are unaffected, and the SU(2)
symmetry is unbroken by the fermion masses, a behaviour we are familiar with
from the special case of pure fundamental fermions. Furthermore, this result
is self-consistent, 
as we have the same minima and thus the same symmetry pattern in the $m
\rightarrow 0$, and $m \rightarrow \infty$
cases.
That is  the fundamental fermions play a more
dominant role than the adjoint ones in determining the symmetry pattern.

\section{Conclusions}

In this work, we have constructed the effective potential for the Wilson loop in the SU(2) gauge
theory with $N_f$ massive fundamental and $N_a$ massive adjoint
fermions on $S^1 \times M^4$  in the one-loop level,
assuming periodic boundary condition for the gauge field,
and the general boundary conditions for fermions
with arbitrary phase,
and investigated the symmetry structure of the vacuum.

Our results can be summarized as follows:

($i$)For the adjoint fermions, the symmetries of
the system depend critically on both  the bulk mass and the bc parameters.  
We have considered the  special limits of the 
general case,
namely the regime of Hosotani with massless fermions and arbitrary
boundary conditions, that of Takenaga with massive fermions and periodic 
boundary conditions ($\delta_a=0$),
and that of Davies and McLachan (the simplest of them all)
with massless fermions, and periodic boundary conditions.
Our predictions are identical to theirs in the corresponding limits.
We have further observed an interesting phenomenon, that for a special 
value of $\delta_a$
(namely, $\delta_a=0.71$, when $N_a=1$)
both broken and unbroken phases coexist for $z_a \leq 0.7$.
Further analysis of this phenomenon is postponed to a future work.

($ii$) Fundamental fermions can never lead
to a spontaneous breakdown of the gauge symmetry irrespective of
the values of the parameters $z_f$,  $\delta_f$,  and $N_f$.

($iii$)  When there are fundamental and adjoint fermions together (with equal
masses), we first note  that in the massless case there are critical values for the boundary condition
parameters $\delta_a$ and $\delta_f$, 
in deciding the absolute minima. However as these are either one of
$\theta_m=0, \pi$,
the symmetry is intact irregardless
of their preferences. Thus there are no critical values for the
bc parameters as far as symmetry breaking pattern is concerned.
We further checked the 
role played by the masses on the symmetry pattern.
We have observed that the minimum values of the effective potential
change with the variations of z, however not as much to change the global minimum
from $\theta_m=0$ to $\theta_m= \pi$, or vice-versa. There is 
a special value of z, $z=8$,
at which all the curves coagulate to the same limiting degenerate
minima (which always happens at $\delta_f=\pi/2$).
However, as the absolute minima are always either one of the
$\theta_m=0$, $\pi$,  the SU(2)
symmetry remains intact independent of the masses, provided that the
boundary condition parameters are chosen within the allowed region
of the massless regime.
It is interesting that the fundamental fermions play a more dominant role
on the gauge symmetry pattern 
than the adjoint ones, when they act together,
as the result is identical to the pure fundamental fermions case.

As explained in the introduction, one immediate application of the
aforementioned compactification is to use $\theta(x)$ as inflaton.
The cosmological data require the inflaton potential to be rather
smooth and inflaton itself to take super-Planckian values. This
necessitates  the extension of field-theoretic description of
Nature into string territory which is hardly acceptable. However,
as already pointed out in~\cite{Arkani-Hamed1,Arkani-Hamed2} and extended to
massive bulk fields in~\cite{Riotto}, the non-integrable phase
$\theta(x)$ is a perfect inflaton candidate due to its shift
symmetry ( as implied by the higher dimensional gauge invariance).
The novelty provided by our analysis is that possible symmetry
breaking parameter domains are identified, and thus, the theory
below $1/R$ might look like either as an Abelian or non-Abelian
theory. In each case, experimentally favoured four-dimensional
gauge coupling $g_4(1/R)\sim 10^{-3}$~\cite{Arkani-Hamed1,Arkani-Hamed2}
experiences different constraints from experimental data at the
weak scale.

Another point which might be of phenomenological importance
concerns the creation of Q balls. Indeed, the four-dimensional
effective theory for the non-integrable phase possesses either and
Abelian or non-Abelian invariance, and in either case its
self-interactions generate lumps of $\theta(x)$ matter in which
all symmetries are broken~\cite{Coleman,Kusenko}. These lumps of matter are
perfect dark matter candidates. Here one notices that such Q-balls
differ from the Kaluza-Klein Q-balls of~\cite{Demir} in that the
latter rests on the inclusion of all Kaluza-Klein modes whereas
the former is based on only $\theta(x)$ which is the zero-mode of
$A_5(x,y)$. The stability as well as further characteristics of
Q-balls of non-integrable phase factor need further analysis of
(\ref{5}).

\vskip 0.5cm
\noindent
We would like to thank Durmu\c{s}
A. Demir for extremely helpful discussions.

\end{document}